\documentclass[letterpaper,structabstract]{aa}
\usepackage{graphicx}
\usepackage{txfonts}

\begin{document}

\def\deg{$^{\rm o}$}

\title{Comparative analysis of the diffuse radio emission in the galaxy clusters A1835, A2029, and Ophiuchus}
\author{  M. Murgia\inst{1,2}
          \and
          F. Govoni\inst{1}
          \and
          M. Markevitch\inst{3}
          \and
          L. Feretti\inst{2}
         \and
          G. Giovannini\inst{2,4}
         \and
          G.B. Taylor\inst{5}
         \and 
          E. Carretti\inst{2}   
          }
\institute{
              INAF - Osservatorio Astronomico di Cagliari,
              Poggio dei Pini, Strada 54, I--09012 Capoterra (CA), Italy 
           \and   
              INAF - Istituto di Radioastronomia, 
              Via Gobetti 101, I--40129 Bologna, Italy
           \and
              Harvard-Smithsonian Center for Astrophysics, 60 Garden Street,
              Cambridge, MA 02138
           \and 
              Dipartimento di Astronomia, 
              Univ. Bologna, Via Ranzani 1, I--40127 Bologna, Italy
          \and
              University of New Mexico, MSC 07 4220, Albuquerque, NM 87131. G.B. Taylor is also an Adjunct Astronomer at the National Radio Astronomy Observatory.
                }
\date{Received; accepted}

\abstract
{}
{We recently performed a study of a sample of relaxed, cooling core
  galaxy clusters with deep Very Large Array observations at 1.4 GHz.
  We find that in the central regions of A1835, A2029, and Ophiuchus
  the dominant radio galaxy is surrounded by a diffuse low-brightness
  radio emission that takes the form of a mini-halo.  Here we present
  the results of the analysis of the extended diffuse radio emission
  in these mini-halos.  }
{In order to investigate the morphological properties of the diffuse 
radio emission in clusters of galaxies we propose to fit their
azimuthally averaged brightness profile with an exponential,
obtaining the central brightness and the e-folding radius from which
the radio emissivity can be calculated.
We investigate the radio properties of the mini-halos in  
A1835, A2029, and Ophiuchus in comparison with the radio properties of a 
representative sample 
of mini-halos and halos already known in the literature.
}
{We find that radio halos can have quite different 
length-scales but their emissivity is remarkably similar from
one halo to the other. In contrast, mini-halos span a wide
range of radio emissivity. Some of them, like the Perseus mini-halos,
 are characterized by a radio emissivity which is more than 100 times greater
than that of radio halos. On the other hand, the new mini-halos
in cooling core clusters analyzed in this work, namely A2029, Ophiuchus, and A1835,
have a radio emissivity which is much more typical of halos in merging clusters rather than
similar to that of the other mini-halos previously known.
}
{}


\offprints{M .Murgia, email m.murgia@ira.inaf.it}
\maketitle


\section{Introduction}
Radio halos are diffuse, low surface brightness, 
steep-spectrum\footnote{We use the convention $S_{\nu}\propto \nu^{-\alpha}$.} sources ($\alpha\gtrsim 1$),
permeating the central regions of galaxy clusters. These radio sources are extended on megaparsec scales 
and are produced by synchrotron radiation of relativistic electrons with energies of
$\simeq 10$ GeV in magnetic fields with $B\simeq 0.5-1\;\mu$G (see e.g. Feretti \& Giovannini 2008, 
Ferrari et al. 2008, and references therein for recent reviews).

Radio halos are typically found in clusters which show significant
evidence for an ongoing merger (e.g. Buote 2001, Govoni et al. 2004). 
It has been proposed that recent cluster mergers may play an 
important role in the re-acceleration of the radio-emitting relativistic particles,
thus providing the energy that powers these extended sources 
(e.g. Brunetti et al. 2001, Petrosian 2001). A major merger event is 
expected to disrupt cooling cores
and create disturbances readily visible in
an X-ray image of the cluster. Therefore,
the merger scenario predicts the absence of large-scale radio
halos in symmetric cooling core clusters.
However, a few cooling core clusters show the presence of
a diffuse synchrotron emission that extends quite far from the dominant radio
galaxy at the cluster center, forming what is called a mini-halo.
These diffuse radio sources are extended
on a moderate scale (typically $\simeq$ 500 kpc) and,
in common with large-scale halos, have a steep spectrum and a very
low surface brightness.

To date about 30 radio halos are known 
(e.g., Giovannini \& Feretti 2000, Bacchi et al. 2003,
Govoni et al. 2001, Venturi et al. 2007, 2008, 
Giovannini et al. in preparation).
Due to their extremely low surface brightness
and large angular extent ($>$10$'$ at a redshift $z\le$0.1)
radio halos are best studied at low
spatial resolution. In fact, several radio halos were detected
by Giovannini et al. (1999)  in 
the NRAO VLA Sky Survey (NVSS; Condon et al. 1998) and by 
Kempner \& Sarazin (2001) in the Westerbork Northern Sky Survey
(WENSS; Rengelink et al. 1997), 
where the relatively large beam of these
surveys provided the necessary sensitivity to large-scale 
emission to spot these elusive sources.

Due to the combination of their small angular
size and the strong radio emission of the central radio galaxy, 
the detection of a mini-halo requires a much higher dynamic range 
and resolution than provided by available surveys,
thus complicating their detection.
As a consequence, our current observational knowledge on
mini-halos is limited to only a handful of well-studied clusters
(e.g., Perseus: Burns et al. 1992; A2390: Bacchi et al. 2003;
RXJ1347.5-1145: Gitti et al. 2007), and their origin and
physical properties are still poorly understood.

To search for new extended diffuse radio emission in relaxed and 
cooling core galaxy clusters, we recently performed a study
of a small sample of clusters with deep Very Large Array observations 
at 1.4 GHz. We find that in the central regions of A1835, A2029, and Ophiuchus
the dominant radio galaxy is surrounded by a diffuse low-brightness radio 
emission that takes the form of a mini-halo (Govoni et al. 2009, in press; 
hereafter Paper I).
Here we present the analysis of these three new
mini-halos and we investigate their radio properties in comparison 
with those of a statistically significant 
sample of mini-halos and halos already known in the literature and 
for which good VLA radio images at 1.4 GHz are available.
In \S 2 we present an analysis of the mini-halos found in A1835, A2029, and 
Ophiuchus. In particular we fit their
azimuthally averaged brightness profile with an exponential. 
Through the study of their averaged brightness profiles, we 
analyze some of their morphological and physical radio properties 
(i.e., length-scale, central brightness, and emissivity).
In \S 3 we compare the properties of the mini-halos in  
A1835, A2029, and Ophiuchus with those of other halos and mini-halos already available in 
the literature. In \S 4 we test the reliability of 
the fitting procedure. The discussion is presented in \S 5 
 and our conclusions are reported in \S 6. 

Throughout this paper we assume a $\Lambda$CDM cosmology with
$H_0$ = 71 km s$^{-1}$Mpc$^{-1}$,
$\Omega_m$ = 0.27, and $\Omega_{\Lambda}$ = 0.73.

\section{Analysis of the mini-halos in A1835, A2029 and Ophiuchus}
In order to investigate the presence of diffuse extended
emission in relaxed systems, we recently examined a small
sample of cooling core clusters with deep Very Large Array
observations at 1.4 GHz. The combination of resolution and sensitivity
of these observations allow us to increase the number of known large
scale diffuse radio emissions in cooling core clusters. Indeed, we
found that in the central regions of A1835, A2029, and Ophiuchus the
dominant radio galaxy is surrounded by a diffuse low-brightness radio
emission that takes the form of a mini-halo. The basic properties of
these clusters are reported in Table \ref{tab1}, while the details of
the observations are described in Paper I.  Here we present a
morphological analysis of these mini-halos. For this purpose, the
radio images have been convolved with a circular Gaussian beam and
corrected for the primary beam attenuation.

\begin{table}
\caption{The mini-halos analyzed in this work and the basic properties of the radio images at 1.4 GHz.}
\begin{center}
\begin{tabular} {ccccc} 
\hline
Cluster   & z &  kpc/\arcsec     & beam & $\sigma$\\
          &   &                  &      & mJy/beam \\
\hline
A1835      &   0.2532    & 3.91 & 53\arcsec$\times$53\arcsec & 0.025  \\
A2029      &   0.0765    & 1.43 & 53\arcsec$\times$53\arcsec & 0.031  \\
Ophiuchus  &   0.028     & 0.55 & 92\arcsec$\times$92\arcsec & 0.16   \\ 
\hline
\multicolumn{5}{l}{\scriptsize Col. 3: Angular to linear size conversion factor;}\\
\multicolumn{5}{l}{\scriptsize Col. 4: FWHM beam of the radio images;}\\ 
\multicolumn{5}{l}{\scriptsize Col. 5: Image sensitivity}\\
\end{tabular}
\label{tab1}
\end{center}
\end{table}

\begin{table*}
\caption{Radio properties of the mini halos derived from the fit procedure }
\begin{center}
\begin{tabular} {ccccccccccc} 
\hline
Cluster& FWHM        & $r_e$    &$r_e$   &  $I_{0_{MH}}$      & $S_{MH}$ & $I_{0_{PS}}$       & $FWHM_{PS}$ & $S_{PS}$ & $\langle J\rangle$ &$\chi^2_{RED}$ \\
       & arcsec    & arcsec &kpc     & $\mu$Jy/arcsec$^2$  & mJy      & $\mu$Jy/arcsec$^2$  &  arcsec   &       mJy&  erg s$^{-1}$Hz$^{-1}$cm$^{-3}$   &          \\ 
\hline\\
\vspace{0.2cm}

A1835  &  53 & $26^{+18}_{-8}$   &$102^{+70}_{-31}$& $1.78^{+4.95}_{-1.32}$&$6.0^{+0.8}_{-0.7}$ & $8.68^{+0.65}_{-1.11}$&$57^{+2}_{-3}$& $32.2^{+0.8}_{-0.7}$& $3.3^{+2.2}_{-1.3}\times 10^{-42}$  &0.03\\ 
\vspace{0.2cm}

A2029  & 53 & $37^{+4}_{-4}$  &$53^{+6}_{-6}$ & $2.75^{+0.92}_{-0.64}$&$18.8^{+1.3}_{-1.1}$& $129^{+7}_{-7}$& $57^{+1}_{-1}$ &$480.0^{+19.5}_{-19.4}$& $5.4^{+1.9}_{-1.3}\times 10^{-42}$   &0.5\\
\vspace{0.2cm}

OPHIUCHUS & 92 & $191^{+23}_{-19}$   &$105^{+13}_{-11}$&$0.58^{+0.08}_{-0.07}$&$106.4^{+10.4}_{-8.9}$& $2.47^{+0.22}_{-0.21}$& $102^{+6}_{-5}$ &$29.0^{+2.0}_{-2.0}$& $4.7^{+0.9}_{-0.8}\times 10^{-43}$   &2.9\\
\hline
\multicolumn{11}{l}{\scriptsize Col. 2: FWHM of the circular Gaussian beam;}\\ 
\multicolumn{11}{l}{\scriptsize Col. 6: Mini-halo radio flux density calculated up to 3$r_e$;}\\
\multicolumn{11}{l}{\scriptsize Col. 10: Average radio emissivity over the volume of a sphere of radius 3$r_{e}$, $k$-corrected with $\alpha=1$.}\\ 
\end{tabular}
\label{tab2}
\end{center}
\end{table*}

\begin{figure*}
\begin{center}
\includegraphics[width=15cm]{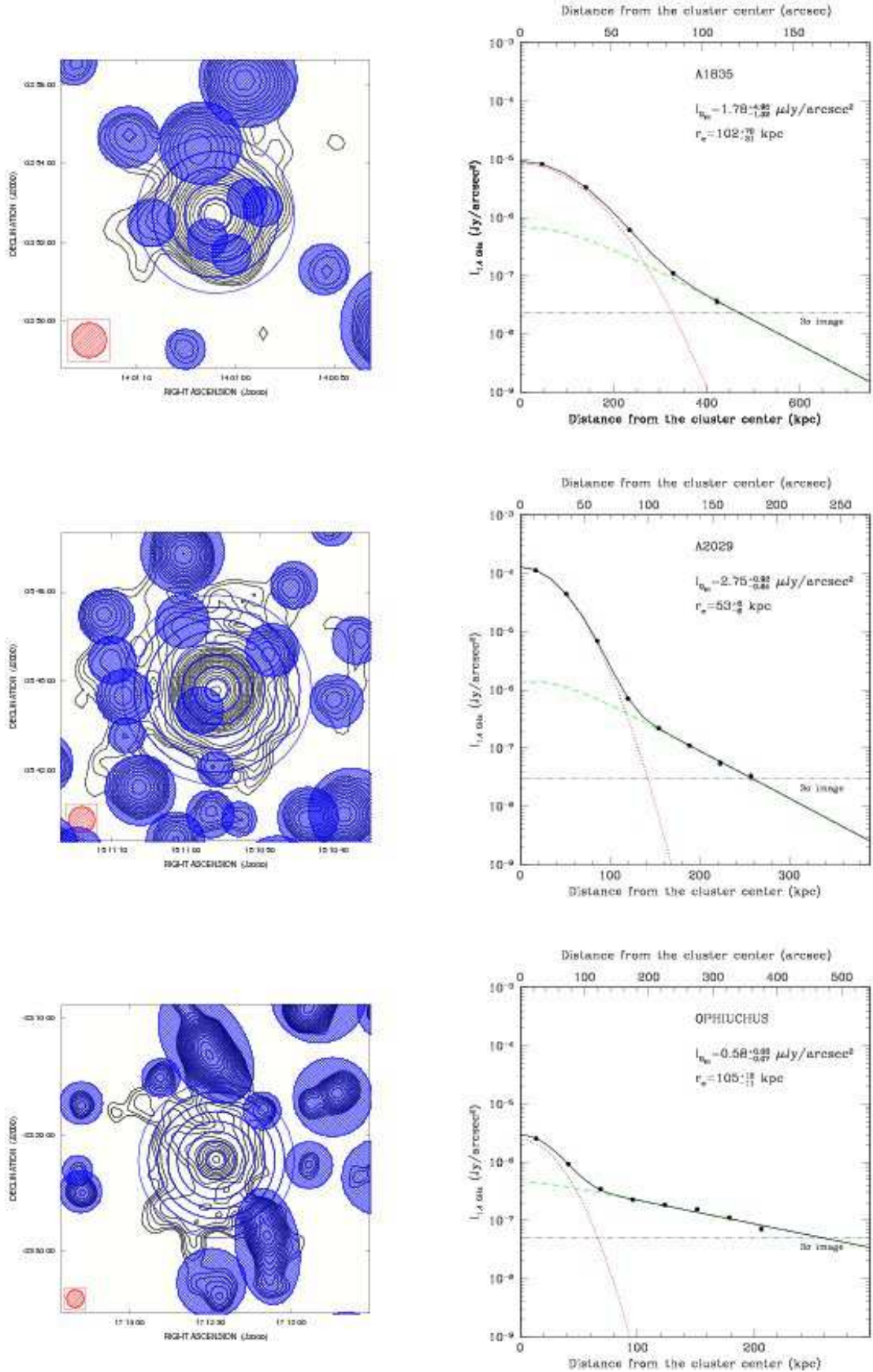}
\caption{
The azimuthally averaged brightness profiles of the radio emission 
in A1835 (top), A2029 (middle), and Ophiuchus (bottom).
The profiles have been calculated in concentric annuli, 
as shown in the left panels.
All the discrete sources (except the central source) 
have been masked.
For each cluster the horizontal dashed-dotted line indicates the 3$\sigma$ noise level of the radio image. 
In our analysis we considered all the data points above the $3\sigma$ noise level. 
The black line indicates the best fit profile 
described by an exponential law (dashed line) representing the mini-halo 
emission, and by a central Gaussian profile (dotted line) representing
the central point source. 
}
\label{fig1}
\end{center}
\end{figure*}

The brightness of halo-like diffuse radio sources in clusters of
galaxies decreases with increasing distance from the cluster
center, eventually falling below the noise level of the radio images. The
signal-to-noise ratio of the VLA images analyzed in Paper I allows us
to trace the mini-halo extension out to a distance from the cluster
center of $\sim$450 kpc, $\sim$250 kpc, and $\sim$200 kpc for A1835,
A2029, and Ophiuchus, respectively (see left-panels of
Fig. \ref{fig1}). However, the size of the diffuse low-surface
brightness emissions as derived from the 3$\sigma$-isophotes seen in
the radio images may be strongly affected by the sensitivity of the
radio observations. It is desirable to define a quantity which
is independent of the signal-to-noise ratio of the radio images in
order to obtain an unbiased estimate of the size of mini-halos and
halos.

Although deviations of the diffuse emission from spherical
symmetry are often observed, the azimuthally averaged radial profiles are 
indeed quite
stable. For this reason, following the same approach used  
by Orr{\`u} et al. (2007), we fit the azimuthally averaged brightness
profiles with an exponential law of the form:
\begin{equation}\label{expfit}
I(r)=I_{0}e^{-r/r_{e}}
\end{equation}
where the two independent parameters
$I_{0}$ and $r_{e}$ are the central radio surface brightness and the e-folding 
radius i.e., the radius at which the brightness drops to $I_{0}/{e}$.
As we will show, the quantity $r_{e}$ can be used to define a length-scale
 which is relatively independent on the sensitivity of the radio images.
The exponential model is attractive in its simplicity and involves a minimal set of free
parameters.

We note that in mini-halos the presence of a strong central radio
galaxy, which appears as a point source at the low resolution of our
observations, complicates the analysis.
Therefore its emission must be known accurately in order
to separate its contribution from that of the mini-halo.

In Fig. \ref{fig1} we present the resulting azimuthally averaged
brightness profiles of A1835, A2029 and Ophiuchus, respectively.  The
profiles have been obtained by averaging the radio brightness in
concentric annuli as shown in the left panels of Fig. \ref{fig1}.  The
annuli are as wide as the half FWHM beam.  In the statistical analysis
all the discrete sources (except the central source) have been
excluded by masking them out.  Moreover, we considered only data
points above the 3-sigma radio noise level as indicated by the horizontal
dashed-dotted line in the plots.

In order to carefully separate the contribution 
of the mini-halo from that of the central radio galaxy,
we fitted the total brightness profiles with a central point source
plus the radio mini-halo diffuse emission
\begin{equation}
 I(r)=I_{PS}(r)+I_{MH}(r).
\label{expodisk+psf}
\end{equation}
In particular the profile of the central point source has been 
characterized by a Gaussian of the form
\begin{equation}\label{expfit}
 I_{PS}(r)=I_{0_{PS}}\,e^{-(r^2/2\sigma_{PS}^2)}
\label{psf}
\end{equation}
while the brightness profile of the mini-halo has been characterized by an 
exponential law of the form
\begin{equation}\label{expfit}
I_{MH}(r)=I_{0_{MH}}\,e^{-r/r_{e}}.
\label{expodisk}
\end{equation}

The resolution of the observations, the masked regions, and the sampling of
the radial profile in annuli, are all factors that can affect the
estimate of the best fit parameters and thus we decided to include
their effects directly in the modelling.

In order to properly take into account the resolution, the
exponential model in Eq.\,\ref{expodisk} is first calculated in a
2-dimensional image with the same pixel size and field of view as that
observed, and then convolved with the same beam by means of a Fast
Fourier Transform. The point-source model in Eq.\ref{psf} is added and
the resulting image is masked exactly in the same regions as for the
observations. Finally, the model is azimuthally averaged with the same
set of annuli used to obtain the observed radial profile. All these
functions are performed at each step during the fit procedure.  As a
result, the values of the central brightness, $I_0$, and e-folding
radius, $r_e$, provided by the fit are deconvolved quantities and
their estimate includes all the uncertainties related to the masked
regions and to the sampling of the radial profile in annuli of finite
width. The fit procedure has been implemented in the software FARADAY
(Murgia et al. 2004).

The best fit model is shown as a continuous black line in the right
panels of Fig. \ref{fig1}.  The mini-halo contribution is indicated by
the green dashed line, while the contribution of the central point
source is indicated with the red dotted line.  The best fit values for
both the mini-halos and the central components are reported in Table
\ref{tab2}, together with their respectively errors and the
corresponding $\chi^2$.

Overall, the observed profiles are well fit by the exponential
model.  In A2029 and Ophiuchus the emission of the mini-halo extends
for several beams from the central point source. It is important to
note the fundamental difference between the method based on the
measurement of $r_{max}$, the mini-halo maximum size from the
3$\sigma$-isophote above the noise\footnote{In this paper, we define
  $r_{max}$ as the distance from the cluster center at which the
  azimuthally averaged brightness profile of the diffuse emission
  reaches the $3\sigma$ noise level of the radio image.}, and the fit
of the e-folding radius, $r_e$. In fact, the mini-halo in A2029 has a
larger extension with respect to that in Ophiuchus: the two mini-halos
have $r_{max}\sim 250$ kpc and $r_{max}\sim 200$ kpc, respectively.
However, the mini-halo emission in A2029 falls off more rapidly and
the corresponding e-folding radius $r_{e}= 53^{+6}_{-6}$ kpc is much
smaller than that in Ophiuchus, $r_{e}= 105^{+13}_{-11}$ kpc.  The fit
is critical for A1835 (see \S 4). This is the most distant cluster of
the three and the emission of the mini-halo is heavily blended with
that of the central point source. The maximum extent for A1835 is
as large as $r_{max}\sim 450$ kpc, while the e-folding radius
obtained by the model fit results, with a large uncertainty, is
$r_{e}= 105^{+70}_{-21}$ kpc.

The central brightness ranges from $I_0 =2.75$ $\mu$Jy/arcsec$^2$ for
A2029 to $I_0=0.58$ $\mu$Jy/arcsec$^2$ for Ophiuchus, indicating a
slightly tendency for the smaller mini-halos to have a higher central
brightness (see \S 3).
In all three cases the best fit model results in a central Gaussian
with $FWHM_{PS}=\sigma_{PS}\times 2.35 \gtrsim FWHM_{beam}$, as
expected for a point like or a slightly resolved source.
From $I_0$ and  $r_e$ we then calculated the flux density $S_{MH}$ (see Table \ref{tab2}) of the mini-halos. 
In particular, the flux density is derived by integrating the brightness profile $I(r)$ up to a radius $r^{\prime}$ from
 the cluster center:
\begin{equation}
 S_{MH}= 2\pi \int_{0}^{r^{\prime}} I(r) r dr=2\pi [1+e^{-r^{\prime}/r_e}(-r^{\prime}/r_e-1)] \cdot r_e^2 I_0 .
\end{equation}

We calculated the flux density of the mini-halos by integrating the surface brightness of the best 
fit exponential model up to $r^{\prime}=3r_e$, giving:
\begin{equation}
S_{MH}= 2\pi f \cdot r_e^2 I_0 ~~~ {\rm (Jy)}
\label{S3re}  
\end{equation}
where $r_e$ and $I_0$ are in units of arcsec and Jy/arcsec$^{2}$, respectively, while $f= (1-4e^{-3})\simeq 0.8$. 
Thus, the flux density in Eq.\ref{S3re} consists of about 
80\% of the total model flux that would be obtained by extrapolating $r^{\prime}$ to $+\infty$. It must be noted that
 the model flux calculated from Eq.\ref{S3re} may differ from the flux density measured by integrating 
 the radio brightness up to the 3$\sigma$-isophote. For instance, in the case of Ophiuchus $r_{max}<3r_e$
 and the model flux of 106 mJy is higher than the flux density of about 80 mJy which is obtained by integrating up 
to the 3$\sigma$-isophote. Instead, in the case of A2029, $r_{max}>3r_e$ and the model flux density of
  18.8 mJy is lower than the flux density of 22.5 mJy obtained by integrating up to the 3$\sigma$-isophote.

In column 9 of Tab.\,2, we also provide the flux density of the
central point source derived from the fit. The flux density of the
central point source, $S_{PS}$, is calculated through:
\begin{equation}
S_{PS}= \frac {2\pi} { 8 \ln(2)} \cdot FWHM_{PS}^2 I_{0_{PS}} ~~~ {\rm (Jy)}
\label{SPS} 
\end{equation}
where $FWHM_{PS}$ is measured in arcsec and $I_{0_{PS}}$ in  Jy/arcsec$^{2}$.
Since mini-halos are optically thin sources,
the radio brightness $I$, at a projected distance $r$ from the cluster centre, is related to the radio emissivity $J(R)$ 
through the line-of-sight integral:
\begin{equation}
 I(r)=\frac{1}{4\pi}\int_{source} J(R)\,dl
\label{intensity}
\end{equation}

where $R=\sqrt{r^{2}+l^2}$.

We calculated a volume-averaged radio emissivity by supposing that all the flux density in Eq.\,\ref{S3re} comes from a 
 sphere of radius 3$r_{e}$:

\begin{equation}
\langle J \rangle \simeq 7.7 \times 10^{-41} (1+z)^{3+\alpha} \cdot \frac{I_0}{r_e}  ~~~~{\rm(erg\,s^{-1}cm^{-3}Hz^{-1})} 
\label{emissivity}
\end{equation}
where $r_e$ and $I_{0}$ are in units of kpc and $\mu$Jy/arcsec$^{2}$, respectively. The factor $(1+z)^{3+\alpha}$ takes
 into account of both the $k$-correction and cosmological dimming of the surface brightness with redshift. 
The radio emissivity for the three mini-halos ($k$-corrected with a spectral index $\alpha=1$) is reported 
 in column 10 of Tab.\,2.

\section{Mini-halos and halos comparison}

Merging and cooling core clusters containing extended diffuse radio sources 
(halos and mini-halos respectively) show different X-ray properties,
indicating a different evolutionary state. A systematic 
and homogeneous comparison between the properties
of extended diffuse radio sources in merging systems with those 
in cooling core clusters, may help in understanding
if the energy necessary to trigger their radio emission may be
related to different physical processes. 
Therefore, we analyzed the azimuthally averaged brightness profiles in a sample
of 12 clusters containing a central radio halo, previously imaged by us.  
Moreover, in order to investigate also a significant number 
of mini-halos, we reduced and analyzed the VLA data at 1.4 GHz of 
the mini-halos in RXJ1347.5-1145 (Gitti et al. 2007), 
A2390 (Bacchi et al. 2003), and Perseus (Pedlar et al. 1990), 
previously reported in the literature.
The cluster list, together with their references is given in Table \ref{tab3}.

\begin{figure*}
\begin{center}
\includegraphics[width=15cm]{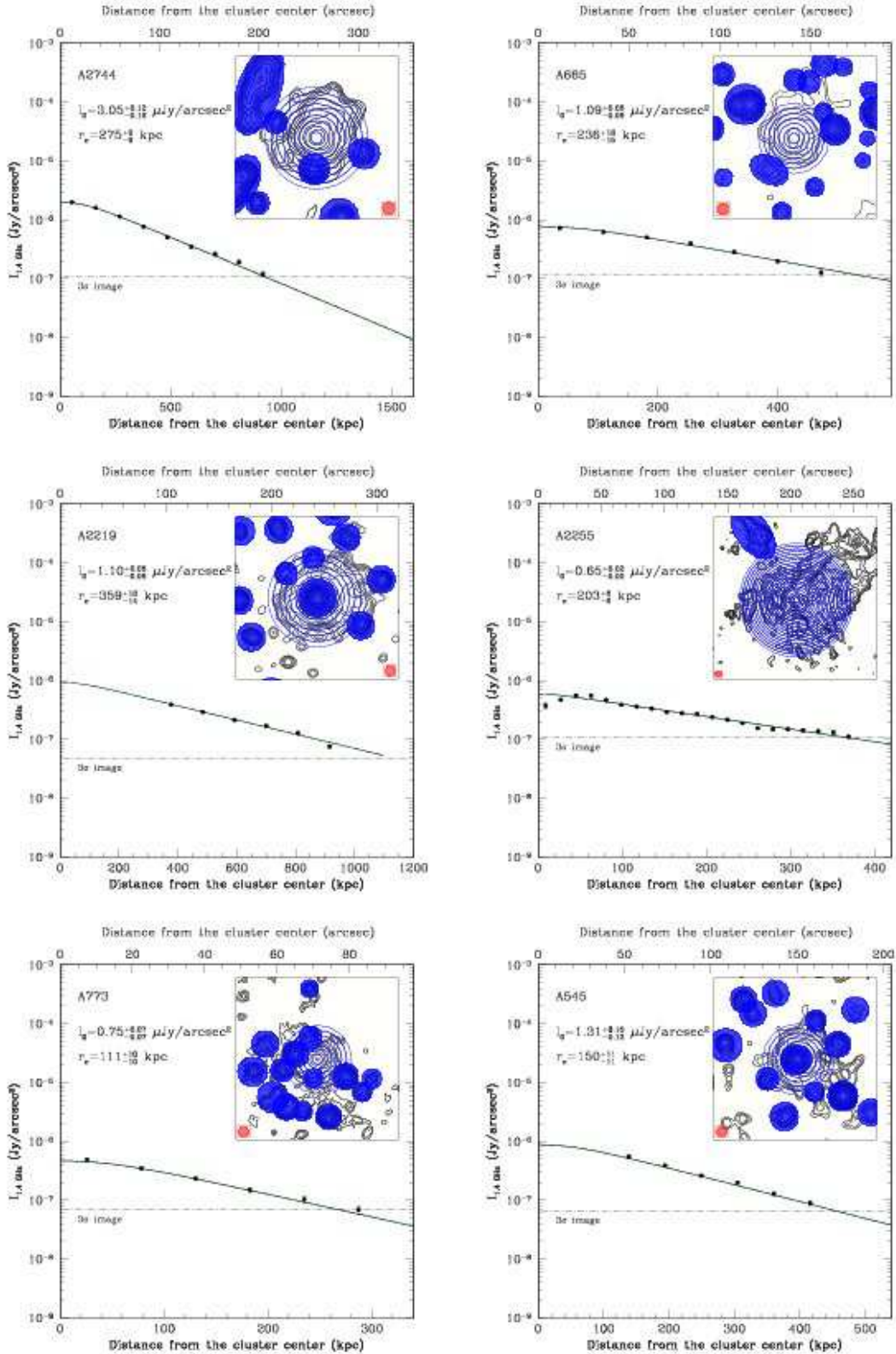}
\caption{
  Azimuthally averaged brightness profiles of the radio halos in A2744
  (Govoni et al. 2001), A665 (Giovannini \& Feretti 2000), A2219
  (Bacchi et al.  2003), A2255 (Govoni et al. 2005), A773 (Govoni et
  al. 2001), A545 (Bacchi et al. 2003).  The profiles have been
  calculated in concentric annuli, as shown in the inset panels.  All
  the discrete sources have been masked.  For each cluster the
  horizontal dashed-dotted line indicates the 3$\sigma$ noise level of
  the radio image.  In the analysis we considered all the data points
  above the 3$\sigma$ radio noise level.  The black line indicates the
  best fit exponential profile.  }
\label{fig2}
\end{center}
\end{figure*}

\begin{figure*}
\begin{center}
\includegraphics[width=15cm]{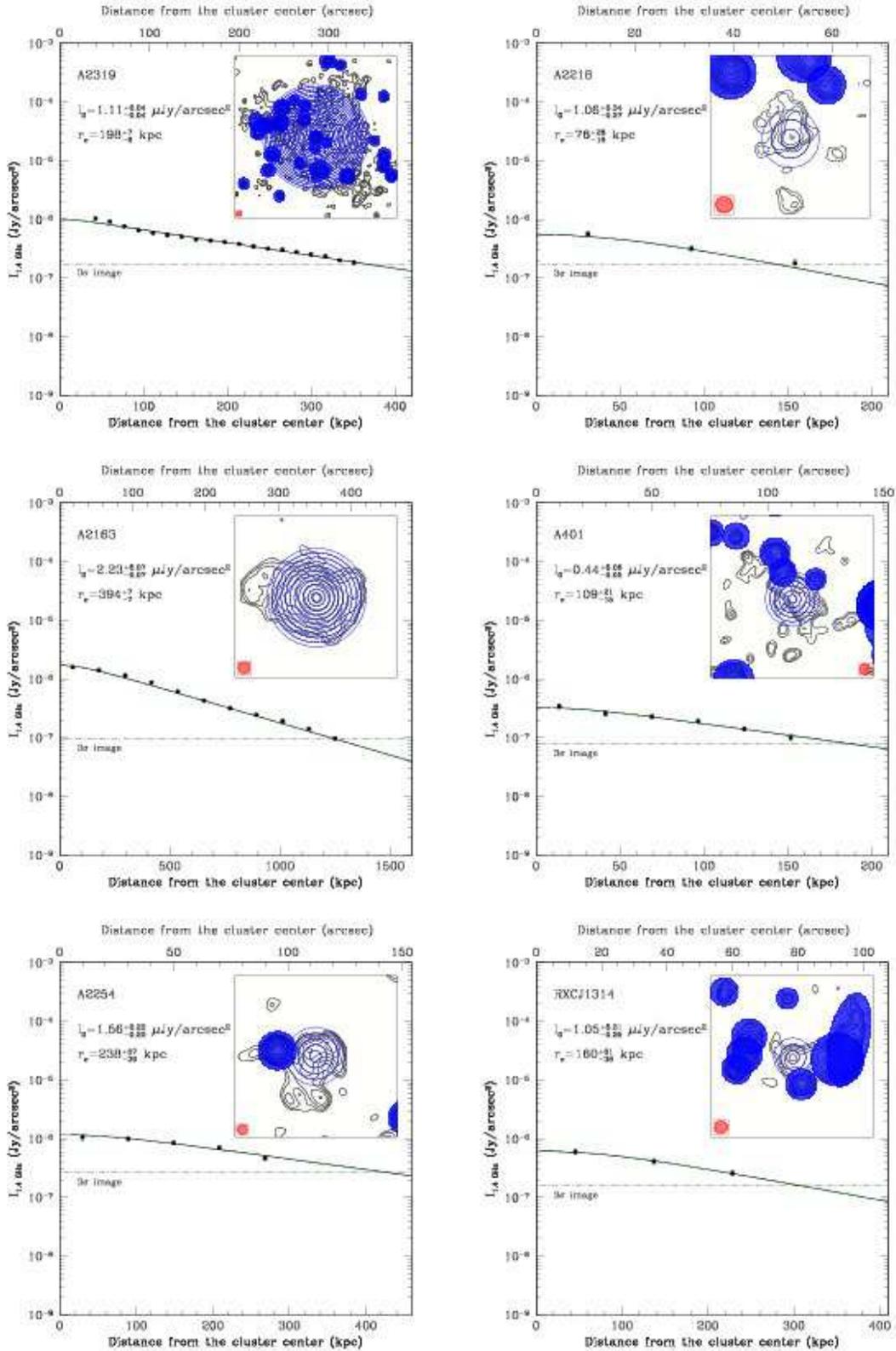}
\caption{
Azimuthally averaged brightness profiles of the radio
halos in A2319 (Feretti et al. 1997), A2218 (Giovannini \& Feretti 2000), A2163 (Feretti et al. 2001), 
A401 (Bacchi et al. 2003), A2254 (Govoni et al. 2001), RXCJ1314.4-2515 (Feretti et al. 2005).  
The profiles have been calculated in concentric annuli, 
as shown in the inset panels.
All the discrete sources have been masked.
For each cluster the horizontal dashed-dotted line indicates the 3$\sigma$ noise level of the radio image. 
In the analysis we considered all the data points above the 3$\sigma$ radio noise level. 
The black line indicates the best fit exponential profile.
}
\label{fig3}
\end{center}
\end{figure*}

\begin{figure*}
\begin{center}
\includegraphics[width=15cm]{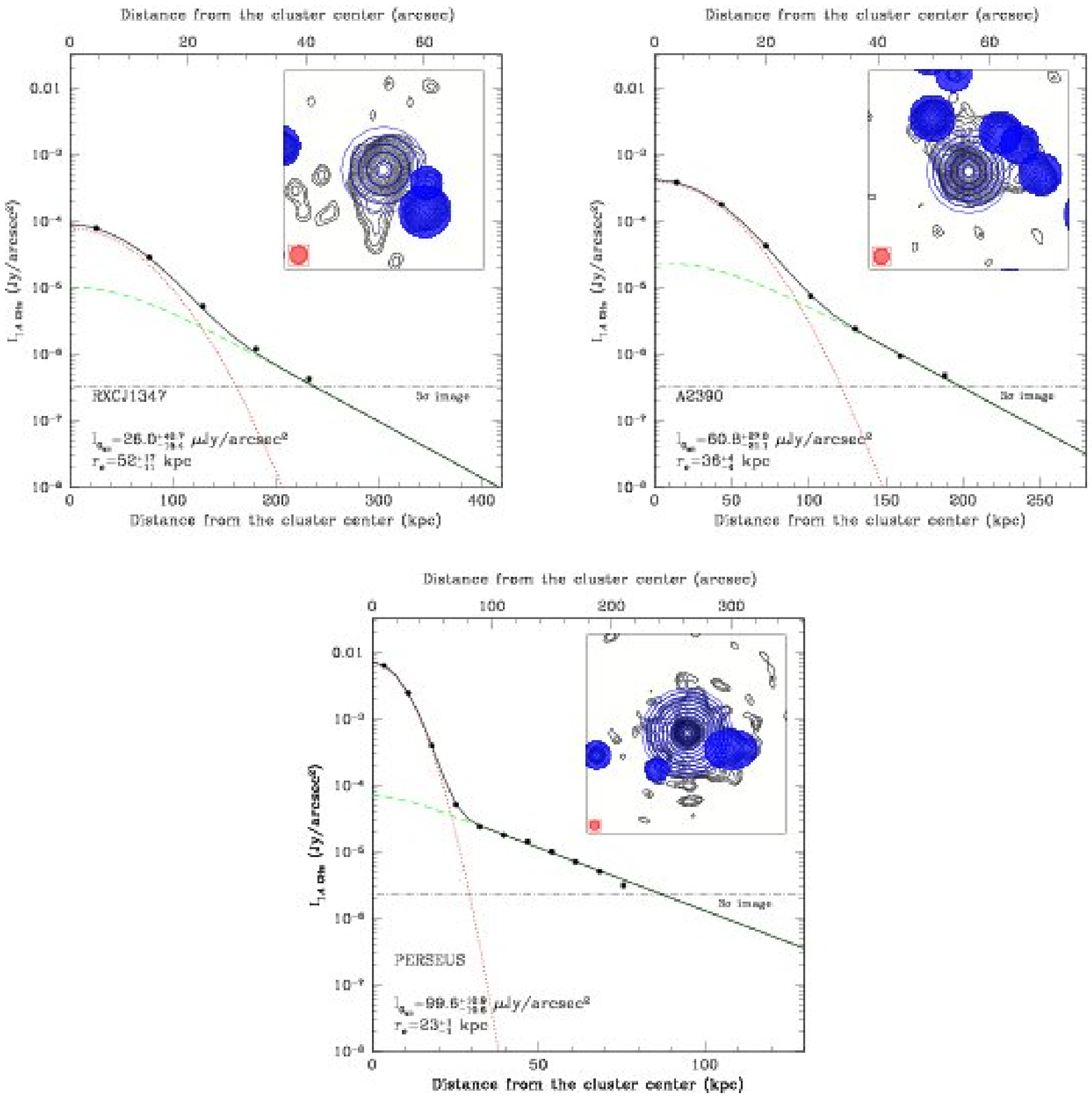}
\caption{
Azimuthally averaged brightness profiles of the mini-halos 
in RXJ1347.5-1145, A2390, and Perseus, previously
imaged in the literature (Gitti et al. 2007, Bacchi et al. 2003, 
Pedlar et al. 1990).
The profiles have been calculated in concentric annuli, 
as shown in the inset panels.
All the discrete sources (except the central one) 
have been masked.
For each cluster the horizontal dashed-dotted line indicates the 3$\sigma$ noise level of the radio image. 
In the analysis we considered all the data points above the 3$\sigma$ radio noise level. 
The black line indicates the best fit profile 
described by an exponential law (dashed line), representing the mini-halo 
emission, and by a central Gaussian profile (dotted line), representing
the central point source. 
}
\label{fig4}
\end{center}
\end{figure*}

Indeed the analysis lacks some well known radio halos and mini-halos, 
however, the number of
clusters analyzed here is representative of these classes of radio
sources and contains both extended (e.g. A2163, A2744) and small halos
(e.g. A401, A2218) in merging clusters and several mini-halos in cooling core 
clusters.
In Fig. \ref{fig2} and \ref{fig3}  we present the resulting azimuthally averaged 
brightness profiles
of the radio halos, while in Fig. \ref{fig4} we present those of the three mini-halos. 
The profiles have been obtained by averaging the radio brightness 
in concentric annuli as shown in the inset of each panel. 

The analysis of the three previously known mini-halos has been
performed with the same approach proposed for A1835, A2029, and
Ophiuchus -- by fitting the brightness profiles with a central
Gaussian plus an exponential law, and by masking all the discrete
sources, except the central one.  The analysis of the halos has been
performed by fitting the brightness profiles with only an exponential
law and by masking all the discrete sources.
As in the case of A1835, A2029, and Ophiuchus the diffuse synchrotron
emission both of the halos and mini-halos are quite well fit by an
exponential law characterized by a central surface brightness $I_0$
and an e-folding radius $r_e$.  The best fit values of the fit
procedure, together with their respectively errors, and the corresponding $\chi^2$, are reported in Tab. 3.

\begin{figure*}
\begin{center}
\includegraphics[width=12cm]{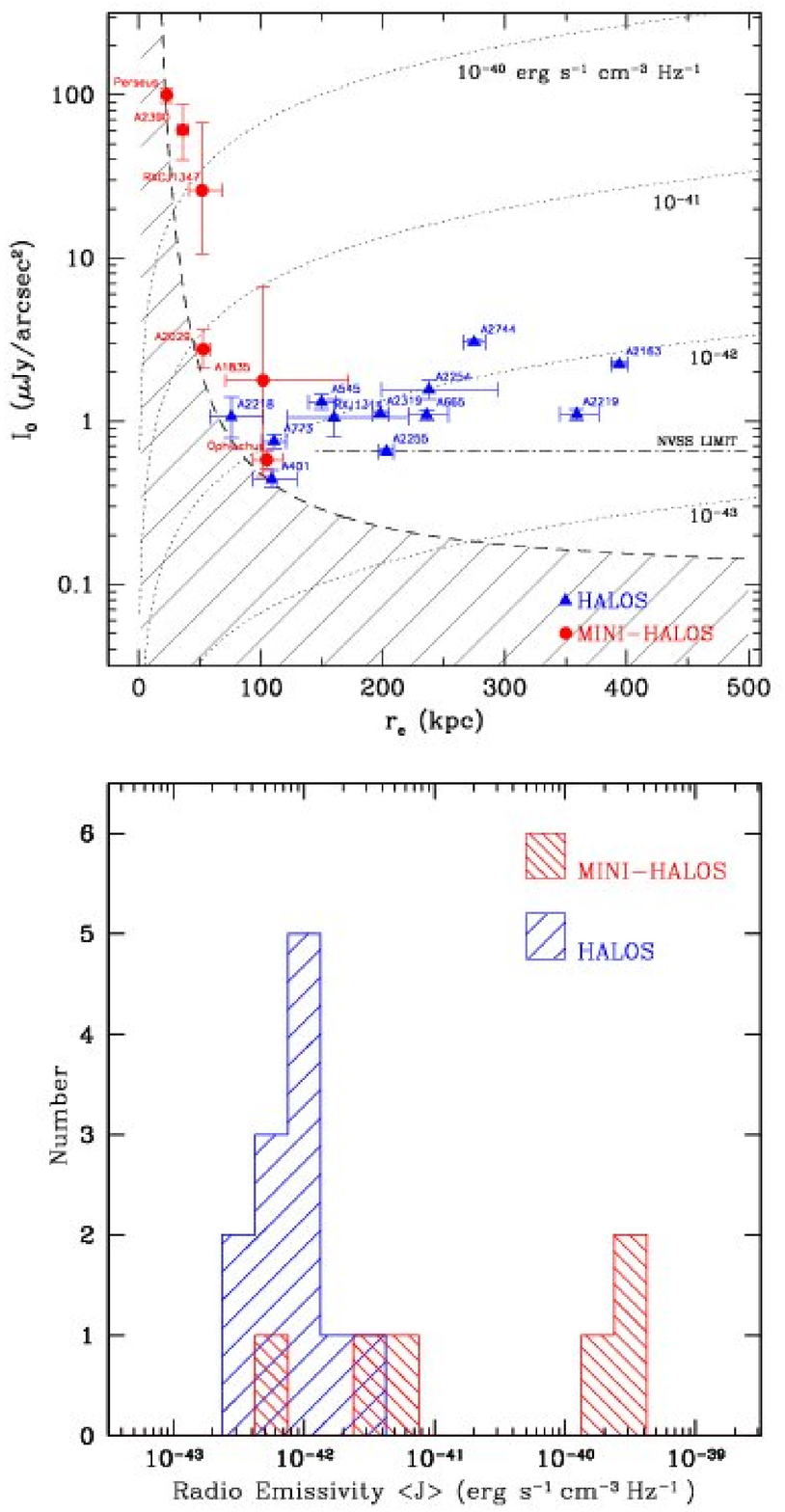}
\caption{
Top: Best fit central brightness $I_0$ ($\mu$Jy/arcsec$^2$) versus
the length-scale $r_e$ (kpc) both for halos (blue triangles) and mini-halos 
(red dots). 
The dotted lines indicate regions of constant emissivity, namely 0.1, 1, 10, 100 times the average 
emissivity of radio halos which is $\langle J \rangle \simeq 10^{-42}$ erg\,s$^{-1}$cm$^{-3}$Hz$^{-1}$. They have been traced 
 from Eq.\ref{emissivity} by assuming  a putative redshift of $z=0.18$ (the average redshift of our sample) and 
 a spectral index $\alpha=1$. The dashed line represents the detection limit expected for a mini-halo or a halo at $z=0.18$ observed in a deep image 
with a beam of $25\arcsec$ and a sensitivity level of 25 $\mu$Jy/beam, see text. The dot-dashed line represents the $3\sigma$ sensitivity
 level of the NVSS. Bottom: Distribution of the $k$-corrected radio emissivities listed in Tables 2 and 3   
both for halos (in blue) and mini-halos (in red).  }
\label{fig5}
\end{center}
\end{figure*}

In the top panel of Fig. \ref{fig5} we show the best fit central
brightness $I_0$ ($\mu$Jy/arcsec$^2$) versus the length-scale $r_e$
(kpc) both for halos (blue triangles) and mini-halos (red dots).  The
length scale $r_e$ in this sample of halos ranges between
$\simeq$70-400 kpc. While the central brightness $I_{0}$ ranges
between $\simeq$0.4 - 3 $\mu$Jy/arcsec$^{2}$.  All the halo sources of
the sample populate a well defined region of the $I_0-r_e$ plane.
There is a tendency for halo sources with a higher central brightness
to have a greater length-scale.  By comparing the location of the halo
sources in the $I_0-r_e$ plane with the dotted lines of constant
emissivity we found, for the first time, that all halos have a
markedly similar radio emissivity \footnote{ However, we cannot
  exclude the possibility that large halos with faint surface
  brightness ($I_{0}<0.5$ $\mu$Jy/arcsec$^{2}$) could have been missed
  by the current searches at 1.4 GHz such that we are just seeing the
  ``tip of the iceberg''. As an example, the $3\sigma$ sensitivity
  level of the NVSS is indicated by the dot-dashed line in the top
  panel of Fig.\,5.  }.

In contrast, the mini-halo distribution in the $I_0-r_e$ plane appears
much more scattered.  In our sample there are no mini-halos with a
$r_e$ larger than $\simeq 100$ kpc.  They have a tendency to populate
the low $r_e$, while the central emissivity can be two orders of
magnitude higher than in halos.  By comparing the location of the
mini-halo sources in the $I_0-r_e$ plane with the lines of constant
emissivity we found that although some mini-halos (like Ophiuchus and
A1835) are quite comparable to the halo sources, in general mini-halos
appear clearly separated from the halo sources, showing a larger spread
in radio emissivity.

As previously noted, there is a tendency for the smaller mini-halos of
our sample to have a higher central brightness. But this trend must be
considered carefully, because a selection effect may be present. The
dashed line in top panel of Fig. \ref{fig5} indicates the minimum
central brightness required for a mini-halo (or a halo) with a given
e-folding radius to be imaged with at least 4 resolution elements
(beams) in a deep image with a sensitivity of $\sigma =25$
$\mu$Jy/beam, and a 25 arcsec beam.  These numbers are appropriate for the
deepest images taken with the VLA at this resolution and frequency
(see e.g. Govoni et al. 2005).  The detection limit is calculated
assuming a putative redshift of $z=0.18$ (the average redshift of the
mini-halos and halos analyzed in this work). In practice, for a
mini-halo (or a halo) to be detected, it is required that $r_{max} >
2\cdot FWHM_{beam}$, where $r_{max}$ is implicitly defined by the
equation: $I_{0}\exp(-r_{max}/r_e)=3\sigma$ which implies $I_{0} \ge 3
\sigma \exp(2\cdot FWHM_{beam}/r_e)$. The shaded region limited by
this detection threshold is hardly accessible by current radio
interferometers at 1.4 GHz.  For instance, an object with $r_{e}\simeq
50$ kpc would be detectable only if brighter than $I_{0} \gtrsim
1\mu$Jy/arcsec$^{2}$. As a result, the observed tendency for the
smaller mini-halos to be brighter must be considered prudently. In any
case, the possible selection effect mentioned above seems not to
influence the result that mini-halos have systematically lower values of
$r_e$ than halos. Mini-halos appear effectively smaller than halos and
this could reflect the higher density of the intergalactic gas in
the central regions of cooling clusters.

For a better comparison between halos and mini-halos in terms of
emissivity, in Fig. \ref{fig5} (bottom panel) we show the distribution
of the radio emissivity (calculated by using Eq.\ref{emissivity}) both
for halos (in blue) and mini-halos (in red).  We find that radio halos
have a surprisingly narrow radio emissivity distribution around an
average value of about $10^{-42}$ erg s$^{-1}$cm$^{-3}$Hz$^{-1}$. In
contrast, mini-halos span a wide range of radio emissivity. Some of
them, like the Perseus mini-halo, would be characterized by a radio
emissivity more than 100 times larger than that of a typical halo,
quite in agreement with the recent finding by Cassano et al. (2008).
We note, however, that the three new mini-halos analyzed in this work
are all characterized by brightness and size very similar to that of the
smaller halos and thus their radio emissivity is comparable to that of
halos in general.

\begin{figure*}
\begin{center}
\includegraphics[width=12cm]{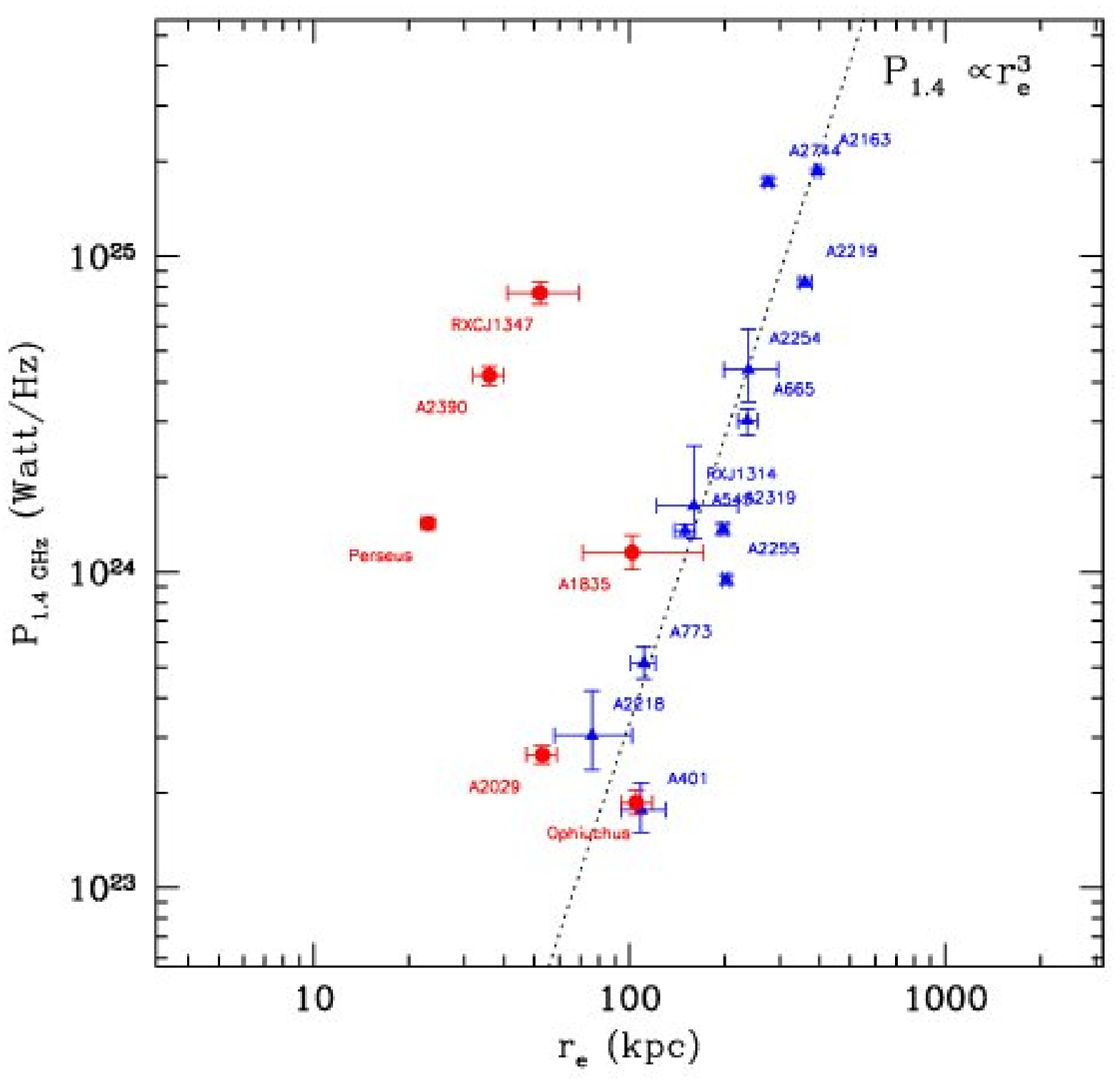}
\caption{Radio power at 1.4 GHz versus e-folding radius in kilosparsecs for mini-halos (red dots) and halos (blue triangles).
The dotted reference line represents
the expected radio power if we assume for the emissivity the average value we found for radio halos 
$\langle J \rangle= 10^{-42}$ erg\,s$^{-1}$cm$^{-3}$Hz$^{-1}$ and for the volume $V$ the volume of a sphere a radius $3r_e$: $\log P_{1.4}\simeq 23.52+3\log(r_{e}/100 {\rm kpc})$.}
\label{fig6}
\end{center}
\end{figure*}

As a further check, we also examined the trend of the radio power
(calculated from the flux density in Eq.\ref{S3re}) versus the
e-folding radius of mini-halo and halos, see Fig. \ref{fig6} . This
plot is similar to that presented in Cassano et al. (2008) except that
they calculated the radius as $\sqrt{r_{min} \times r_{max}}$,
$r_{min}$ and $r_{max}$ being the minimum and the maximum radius
measured using the 3$\sigma$ radio isophotes.  As pointed out by Cassano
et al. (2007), the powerful radio halos are the more extended sources
while the fainter ones are smaller in size. Fig. \ref{fig6} shows
indeed a tight correlation between the radio power and the e-folding
radius of radio halos. We find that the radio power increases as
$P_{1.4}\propto r_{e}^{3}$. In our analysis, the slope of the
correlation is in agreement with the finding that the radio emissivity
has a very small scatter among radio halos. In fact, the radio power
scales as $P_{1.4}= J\cdot V$, where $V$ is the source volume and $J$
is the radio emissivity. If we assume that the emissivity does not
change from one halo to the other it follows that $P_{1.4}\propto
r_{e}^{3}$.  The dotted reference line in Fig. \ref{fig6} represents
the expected radio power if we assume for the emissivity the average
value we found for radio halos $\langle J \rangle\simeq 10^{-42}$ erg
s$^{-1}$cm$^{-3}$Hz$^{-1}$ and for the volume $V$ the volume of a
sphere a radius $3r_e$: $\log P_{1.4}\simeq 23.52+3\log(r_{e}/100 {\rm
  kpc})$.

We note that Cassano et al. (2008) found also a correlation between
the power and the size of mini-halos. In fact, the three mini-halos we
have in common (A2390, RXCJ1347, and Perseus) all seem to be aligned in a
similar but offset correlation with respect to radio halos. However,
with the addition of the new objects found in Paper I and analyzed
here, we cannot define any clear correlation for the mini-halos, in
agreement with the large spread in emissivity shown in Fig.
\ref{fig5}.  The mini-halos in A1835, Ophiuchus, and A2029 appear
clearly separated from the other mini-halos although this separation
into two distinct groups is likely just due to the small-number
statistics.

\begin{figure*}
\begin{center}
\includegraphics[width=12cm]{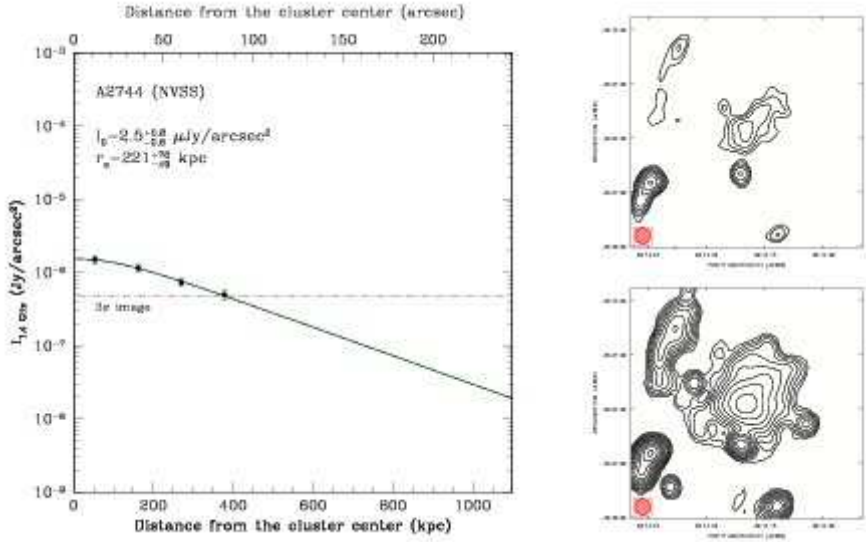}
\caption{
Left: fit of the azimuthally averaged brightness profiles 
calculated from the NVSS image of the cluster A2744.
Right: Contours of the NVSS image in comparison with the
contours of the deeper VLA image for A2744 (Govoni et al. 2001).
In both images the first contour is drawn at 3$\sigma$ and the rest 
are spaced by a factor $\sqrt{2}$.
The sensitivity (1$\sigma$) of the NVSS is 0.45 mJy/beam while the 
sensitivity of the deeper image is 0.1 mJy/beam.
The radius as derived from the contour levels depends on the sensitivity 
of the radio image, while the fit procedure give similar results in 
images at different sensitivity.
}
\label{fig7}
\end{center}
\end{figure*}

As a final remark, we stress once again that our analysis should be
relatively independent of the noise level of the radio images.  To
ensure that our approach is not dependent on the sensitivity of the
radio observation we retrieved for A2744 (one of the clusters analyzed
here) the NVSS image.  We derived the azimuthally averaged brightness
profile as computed for the deeper observations (same rings and
mask), and we then fit the NVSS profile with an exponential law.
The resulting fit is shown in the left panel of Fig. \ref{fig7}.  The
contours of the NVSS image are shown on the top right panel of
Fig. \ref{fig7} in comparison with the contours of the deeper VLA
images on the bottom right panel.  As expected, the radius as derived
from the contour levels yields very different values in images with
different sensitivities.  One should consider that the same effect
may appear in images with comparable sensitivity of objects
characterized by the same intrinsic size but different
brightness. Fainter halos (or mini-halos) tend to ``sink'' into the
noise appearing smaller than they really are.  On the other hand, in
comparing the fitting results of the NVSS profiles in Fig. \ref{fig7}
with the fitting results of the deeper images in Tab. \ref{tab3}, we
find the results to be in good agreement.  Therefore we are confident
that the method discussed here, based on the determination of
length-scale and central brightness in extended structures, is less
subject to the S/N ratio of the radio images and thus better suited
for comparative studies of the general properties of diffuse halo-like
sources.

\section{Monte Carlo statistical analysis of the fitting procedure}

The fit of the exponential model described in \S 2 could be critical
in the case of those mini-halos in which the e-folding radius is
smaller than the FWHM of the central point source. The mini-halos in
A1835, A2029, RXCJ1347, and A2390 are all in this category.  However,
it should be noted that not only the observing resolution but also the
brightness of the mini-halo is crucial for the determination of the
e-folding radius. If the mini-halo is bright enough, the exponential
profile will emerge from the central point source and it will be
observable out to a large distance from the cluster center, making the
fit of the e-folding radius $r_e$ possible.  The example of A2029 is
representative of this situation. The e-folding radius derived from
the fit, $r_e=37\arcsec\pm4 \arcsec$, is below the FWHM beam of the
observation which is 53\arcsec. Nevertheless, the relative intensity
of the mini-halo with respect to the noise permits the fit of the
exponential model from the external part of the mini-halo profile that
does not overlap with the central point source. The limiting cases of
A1835 and RXCJ1347 (the two more distant mini-halos in our sample) are
more critical and the behaviour of the fitting procedure must be
carefully checked.

\begin{figure*}
\begin{center}
\includegraphics[width=18cm]{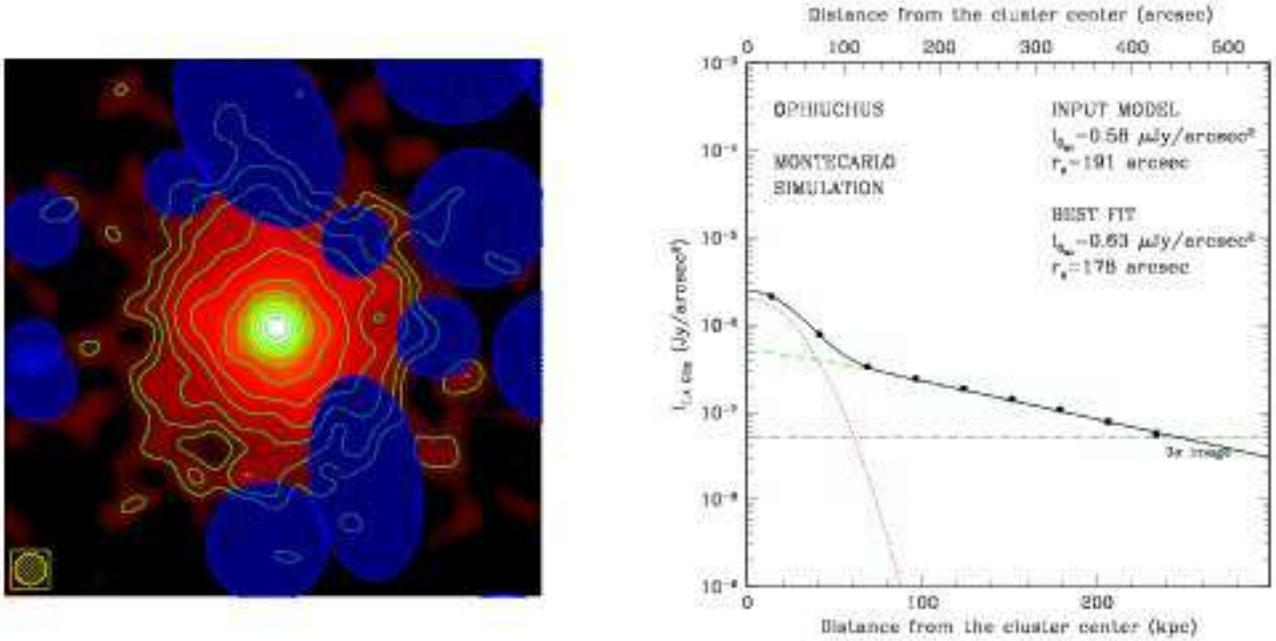}
\caption{
Synthetic image (left panel) and fit profile (right panel) drawn from the Monte Carlo simulation of the Ophiuchus mini-halo.}
\label{OPHIUCHUS_MC}
\end{center}
\end{figure*}

\begin{figure*}
\begin{center}
\includegraphics[width=18cm]{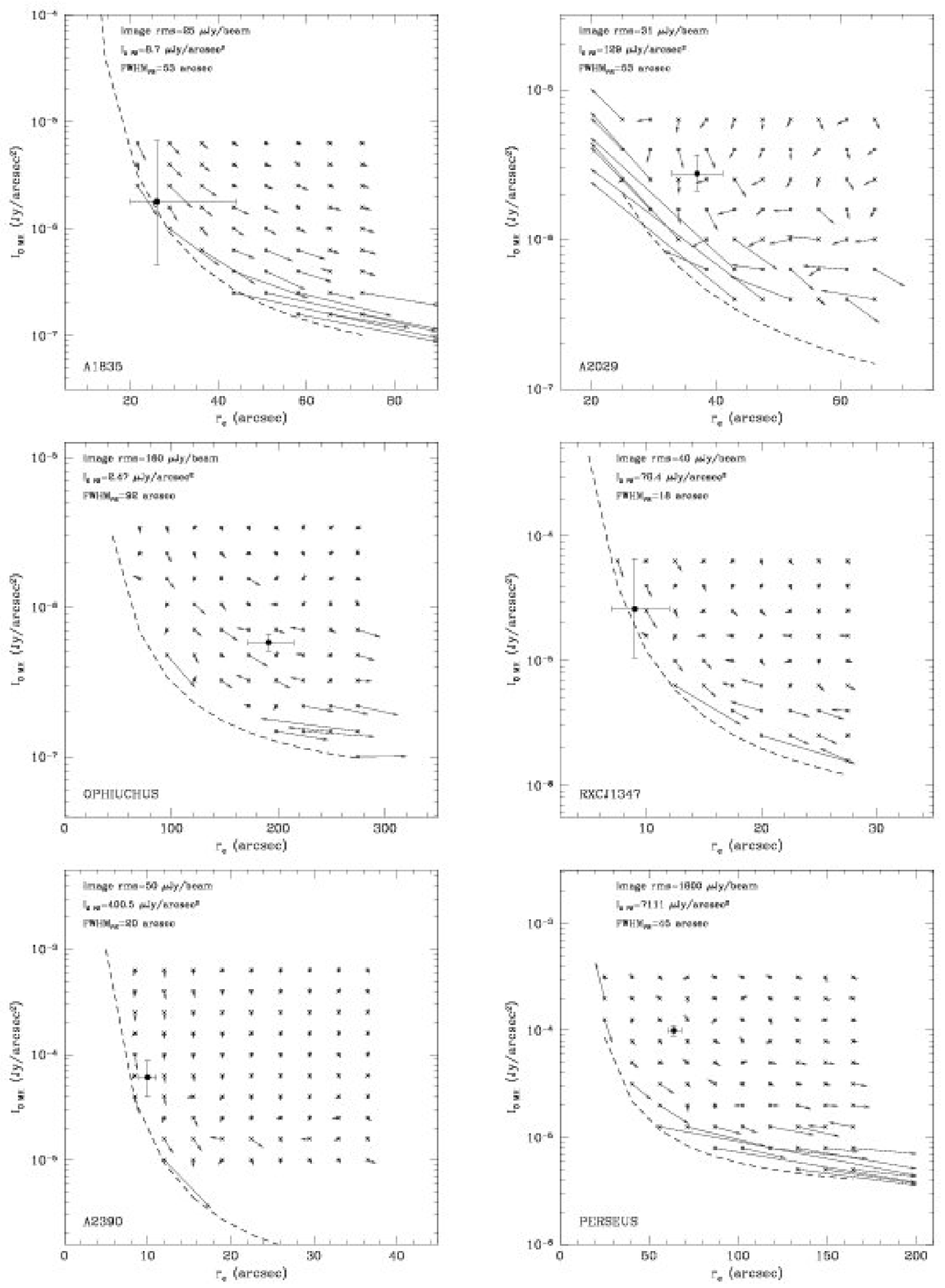}
\caption{
Monte Carlo simulations for the six mini-halos analyzed in this work, see 
text for further details.}
\label{MC}
\end{center}
\end{figure*}

In order to assess the reliability of the best fit parameters, we
created a set of Monte Carlo simulations with the aid of the
Cybersar-OAC computer cluster. For each mini-halo we varied the values
of $I_{0}$ and $r_{e}$ in a regular grid that contained the best fit
parameters reported in Tab. \ref{tab2} and Tab. \ref{tab3}.  For each
pair of $I_{0}$ and $r_{e}$ values, we created a synthetic image with
the same field of view, pixel size, FWHM beam, noise level, and masked
regions as that observed. A central point source with the same
intensity as the observed one is added to the exponential model. An
example of such image, drawn from the Monte Carlo simulation of the
Ophiuchus mini-halo, is presented in left panel of
Fig. \ref{OPHIUCHUS_MC}. We then feed the synthetic image into the fit
procedure and check for any systematic difference between the input
and the output values of $I_{0}$ and $r_{e}$.  In the right panel of
Fig. \ref{OPHIUCHUS_MC} we show the fit obtained from the synthetic
image on the left. We can see that in the case of the Ophiuchus
mini-halo, which is particularly extended, the difference between
input and the output values of $I_{0}$ and $r_{e}$ is fully consistent
with the uncertainties reported in Table \ref{tab2}.  In Fig. \ref{MC}
we present the result of the Monte Carlo simulations for all six
mini-halos analyzed in this work. For each mini-halo we plot the grid
of the input values of $I_{0}$ and $r_{e}$ (crosses) with the relative
arrows representing the displacement of the output parameters from the
fit procedure.  The dashed line represents the detection limit for
each mini-halo and has been traced with the same criterion as in top
panel of Fig. \ref{fig5}. Here the exact values for the FWHM beam and
noise level of the specific observations have been used. In each
panel, the dot indicates the values of the actual best fit parameters
with their uncertainties.  Clearly, for bright and extended
mini-halos, the fit procedure is able to recover the original
parameters with high precisions. Obviously, near the detection limit
the situation can be much worse and large systematic effects are
observed. However, we note that the fit procedure is able to recover
the input e-folding even if it is significantly smaller then the FWHM
beam, provided that the value of $I_0$ is sufficiently high (see
e.g. the case of A2390 with $r_e\simeq 10\arcsec$ and $I_0 >10$
$\mu$Jy/beam).

In summary, the results of the Monte Carlo simulation tell us that the
fit procedure is able to constraint the model parameters with accuracy
and without systematic effects in the cases of Ophiuchus, Perseus,
A2029, and A2390. The mini-halos in A1835 and RXCJ1347 are very close
to the detection limit and the best fit parameters could be affected
by a systematic bias.  In particular, they could be smaller and
brighter than what is inferred by the fit to the exponential model,
although the bias appears to be comparable to the reported
uncertainties.

\section{Discussion}

The exponential model fitting method provides an estimate of the
brightness and size of the diffuse, halo-like, radio emission in
cluster of galaxies using a minimal set of free parameters.  In this
section we discuss the exponential profile in comparison with
alternative models which may have a much clearer physical meaning, but
require the introduction of a number of additional assumptions and/or
free parameters.

One main difficulty in the study of the non-thermal radio halos and
mini-halos in clusters of galaxies is that the synchrotron emissivity
essentially traces the product of electron and magnetic field energy
densities so disentangling the two contributions is not easy. Another
source of uncertainty is related to the effective shape of the energy
spectrum of the relativistic electrons across the cluster which is
poorly known.  Given their comparatively short radiative life-times,
the synchrotron electrons cannot diffuse into the large volume of
space involved but rather they must be either continuously injected
and/or re-accelerated in situ throughout the intra-cluster medium
(e.g., Dennison 1980, Jaffe 1977, Dolag \& Ensslin 2000; Brunetti
et. al 2001).  The spectral energy distributions for the cosmic ray
electrons predicted by these models can be quite different, however
from the current data we are not able to discriminate between the two
processes definitively.  Our intent here, is to highlight the role of
the magnetic field in determining the radial profile of the radio
emission.  For illustrative purposes, in the following we discuss
simplified models for the distribution of particles and magnetic
fields in clusters that are often discussed in literature and we
compare them with the exponential profile.
 
\subsection{Injection models}

We can compare a continuous injection of cosmic ray electrons in the 
intra-cluster medium with a power law distribution in the energy range from 
 $\gamma_{min}m_ec^2$ to $\gamma_{max}m_ec^2$ ($\gamma_{max}\gg \gamma_{min}$). In the stationary case, and for an 
emission spectrum with index $\alpha=1$, from standard synchrotron formulas (Rybicki \& Lightman 1979) we obtain for 
the radio emissivity:

\begin{equation}
J_{\nu}=2.21\times 10^{-10}  \frac{Q_{el}} {B_{\mu G}^{2}+B_{CMB}^2} B_{\mu G}^{2}\nu_{GHz}^{-1} ~~~({\rm erg\,s^{-1}cm^{-3}Hz^{-1}})
\label{CI}
\end{equation}
where $B_{CMB}=3.25(1+z)^2~\mu$G is the equivalent magnetic field associated with the cosmic microwave background while $Q_{el}$ is the production rate of relativistic electrons in units of erg s$^{-1}$cm$^{-3}$. In Eq.\,\ref{CI} we supposed that the electron population is isotropic and we average over all the possible directions between the magnetic field and the line-of-sight, i.e. the field is completely tangled. 

The energy density of the cosmic ray electrons can be calculated as

\begin{equation}
u_{el}\simeq 7.74 \times 10^{20}\frac{Q_{el}} {B_{\mu G}^{2}+B_{CMB}^2} \gamma_{min}^{-1}  ~~~~~({\rm erg\, cm^{-3}}).
\label{uel}
\end{equation}

In the case of a perfect equipartition between the energy densities of particles and fields ($u_{el}= B^2/8\pi$), we have that the radio 
source is near minimum energy conditions and the radio emissivity at 1.4 GHz is given by: 

\begin{equation}
J_{1.4}\simeq 8.12\times 10^{-43} \gamma_{min 100} B_{\mu G}^{4} ~~~({\rm erg\,s^{-1}cm^{-3}Hz^{-1}}).
\label{Jsynchro}
\end{equation}
The synchrotron emissivity is indeed very close to the observed value of $\langle J \rangle\simeq 10^{-42}$  erg s$^{-1}$cm$^{-3}$Hz$^{-1}$ in the case of radio halos for a low energy cut-off of $\gamma_{min}$=100 and an average magnetic field strength of about 1 $\mu$G (changing $\gamma_{min}$ from 10 to 1000 causes $B$ to vary from 1.8 to 0.5 $\mu$G).
Assuming a redshift of z=0.18 (the average distance of the radio halos in our sample), from Eq.\,\ref{uel} we find that the required energy supply to explain the 
observed radio halos emissivity in a minimum energy condition is $Q_{el}\simeq 1.1\times 10^{-31}$  erg\,s$^{-1}$cm$^{-3}$. Note that this power is about an order 
of magnitude larger than the total synchrotron luminosity and one order of magnitude less than the thermal X-ray luminosity.

The equipartition magnetic field strength quoted above should be regarded as volume averages. However, there are observational and theoretical arguments to support the idea that inside a cluster the magnetic field strength scales as a function of the thermal gas density, $n_{th}$, so:

\begin{equation}
B(R)=B_0 \left[\frac{n_{th}(R)}{n_{th}(0)}\right ]^{\eta}
\label{eta}
\end{equation}

where the index $\eta$ is expected to be in the range $0.3 - 1$  (e.g., Dolag et al. 2001, Dolag et al. 2002, Murgia et al. 2004, Guidetti et al. 2008, Battaglia et al. 2009).

If the density profile of the thermal gas is described by the standard $\beta$-model (Cavaliere \& Fusco-Femiano 1976), from Eq.\,\ref{Jsynchro} we have

\begin{equation}
J_{1.4}(R)=J_0 (1+R^{2}/r^{2}_c)^{-6\beta\eta}
\label{beta}
\end{equation}

where $r_c$ is the cluster's core radius. The radio  brightness profile is:

\begin{equation}
I_{1.4}(r)=I_0 (1+r^{2}/r^{2}_c)^{-6\beta\eta+0.5}
\label{profking}
\end{equation}
where $r$ is the projected distance from the cluster center.
It is worthwhile to note that the ratio between the radio and X-ray  brightness is proportional to

\begin{equation}
\frac {I_{1.4}}{I_X}\propto T^{-1/2}(1+r^{2}/r^{2}_c)^{-3\beta (2\eta-1)}
\label{radioX}
\end{equation}

where $T$ is the gas temperature. Thus, if $\eta=0.5$, a linear relation is expected between the radio and the X-ray brightness for 
 a perfectly isothermal $\beta$-model.

We now reconsider the radio halos in A2744 and A665, for which an average spectral index $\alpha\simeq 1$ has been determined from 
VLA radio images at 327 MHz and 1.4 GHz by Orr\`u et al. (2007) and Feretti et al. (2004), respectively.
In the top panels of Fig.\,10, we trace the equipartition model profiles by converting to our cosmology the $\beta$-model 
parameters reported by Kempner \& David (2004) and Birkinshaw et al. (1991) for A2744 ($r_{c}=382$ kpc and $\beta=0.79$) and 
A665 ($r_{c}=291$ kpc and $\beta=0.66$), respectively. The equipartition model is very sensitive to the radial profile of the
 magnetic field strength. In order to show this behaviour, we fix $I_{0}$ in Eq.\ref{profking} and choose three particular values for $\eta$. A 
constant field, $\eta=0$, results in a profile which is too flat with respect to the data while 
a magnetic field whose strength decreases with the distance as the gas density, $\eta=1$, results in
 a profile which is too steep. A very good agreement with the data is found for  $\eta\simeq 0.4- 0.5$ for both radio halos, 
implying a magnetic fields whose energy density is proportional to that of the thermal gas. Indeed, in A2744 a linear correlation
 between the radio and X-ray surface brightness has been observed (Govoni et al. 2001). Moreover, a comparison of the equipartition model with $\eta=0.5$ and the exponential fit shows that 
the two profiles are strikingly similar within one megaparsec from the cluster center. The differences between the two profiles
 become appreciable only at much larger distances, where the radio halo emission has fallen below the noise level. 

 In the bottom panels of Fig.\,10 we show the exponential and
 equipartition profiles for the mini-halos in the Ophiuchus and
 Perseus clusters. We adopt a core radius of $r_c=199$ kpc and
 $\beta=0.747$ for Ophiuchus, and $r_c=44$ kpc and $\beta=0.54$ for
 Perseus (Chen et al 2007).  We note that the equipartition profile in
 Eq.\ref{profking} should be considered only outside the cool core
 where the $\beta$-model still provides a reasonable description of
 the gas density.  Moreover, Ophiuchus lacks any spectral index
 information while the mini-halo in Perseus has an average spectral
 index which is slightly steeper than what we have assumed here
 ($\alpha\simeq 1.2$; Gitti et al. 2004). Apart from these caveats,
 however, the exponential and the equipartition profiles are
 remarkably similar in both mini-halos. As found for the radio halos
 in A2744 and A665, the mini-halo in Ophiuchus is well described by an
 equipartition model with $\eta\simeq 0.5$.  In the case of the
 Perseus mini-halo, however, we found $\eta\simeq 0.8$, which implies
 a particularly steep magnetic field radial profile (this result does
 not change significantly by assuming $\alpha = 1.2$ instead of
 $\alpha=1$). Indeed, Burns et al. (1992) suggested that the absence
 of a cluster-wide magnetic field inhibits a large scale halo in
 Perseus.

 We conclude that the exponential profile is very close to the
 expectations of a simple equipartition model if the magnetic field
 energy density roughly scales as the thermal gas density. In the
 framework of the equipartition model, the physical meaning of the
 $3r_{e}$ length-scale obtained by the exponential fit is that it
 marks the point at which the magnetic field strength is decreased to
 about half the value at the cluster center (see bottom panel of
 Fig.\,10). Hence, mini-halos would appear smaller because the
 magnetic field falls off more rapidly with radius.

 It is worthwhile to note that if $B < B_{CMB}$ and $\eta\simeq 0.5$
 (as for A2744, A665 and Ophiuchus), the equipartition condition
 implies that the injection rate should scale roughly as the thermal
 gas density $Q_{el}\propto n_{th}$.  In models that consider a
 continuous production of secondary electrons by hadronic collisions
 in the intra-cluster medium (Dennison 1980), the injection rate of
 cosmic ray electrons is proportional to the product of the densities
 of the gas and cosmic ray protons, $Q_{el}\propto n_{th}
 n_{CRp}$. Usually, it is assumed that $n_{CRp}\propto n_{th}$ and as
 a results $Q_{el}\propto n_{th}^{2}$. As a consequence secondary
 models predict brightness profiles that are generally too steep to
 explain observed trends in radio halos. However, the radio brightness
 profiles in the Perseus mini-halo is steeper and it can be described
 by secondary models as well (Pfrommer \& Ensslin 2004).

\subsection{Re-acceleration models}
The observed radio profiles may be also described in the framework of the re-acceleration scenario (e.g. Gitti et al. 2004). An over-simplified 
re-acceleration model may be considered in which the relativistic electrons energy spectrum is essentially peaked at a characteristic energy, $\gamma_{max}mc^2$, at which 
the re-acceleration gains balance the radiative losses. 
In the stationary case, the energy of the cosmic ray electrons can be calculated as:

\begin{equation}
u_{el}\simeq 7.74 \times 10^{20}\frac{Q_{el}} {B_{\mu G}^{2}+B_{CMB}^2} \gamma_{max}^{-1}  ~~~~~({\rm erg\, cm^{-3}})
\label{uelreac}
\end{equation}
where $Q_{el}=\dot u_{el}$ is the energy supply to the re-acceleration process in units of erg s$^{-1}$cm$^{-3}$.

The emission spectrum will extend up to a maximum frequency of $\nu_{max}=4.2\times10^{-9}B_{\perp\mu G} \gamma_{max}^{2}$ GHz,
beyond which it cuts off exponentially. The synchrotron emissivity in c.g.s. units is:

\begin{equation}
J_{\nu}=2.74\times 10^{-10}  \frac{Q_{el}} {B_{\mu G}^{2}+B_{CMB}^2} B_{\mu G}^{2}\nu_{max}^{-1} F(\nu/\nu_{max})
\label{Reacc}
\end{equation}

where $F(\nu/\nu_{max})$ is the usual synchrotron kernel (Pacholczky 1970) while $\nu$ and $\nu_{max}$ are expressed in GHz. 

In order to reproduce a spectral index $\alpha_{325 MHz}^{1.4 GHz}=1$ between 325 MHz and 1.4 GHz, 
it should be $\nu_{max}\simeq 0.56\cdot(1+z)$\,GHz. At a frequency of $\nu=1.4$\,GHz and at $z=0.18$, the average redshift of the
 sources in our sample, we can calculate $F(\nu/\nu_{max})\simeq 0.27$ (a steeper spectrum with $\alpha_{325 MHz}^{1.4 GHz}=2$ would 
imply $\nu_{max}\simeq 0.31\cdot(1+z)$\,GHz and $F(\nu/\nu_{max})\simeq 0.06$) . 

In the case of a perfect equipartition between the energy densities of particles and fields, $u_{el}= B^2/8\pi$, we find that the radio  source is near the minimum energy condition 
and the radio emissivity at 1.4 GHz is: 

\begin{equation}
J_{1.4}\simeq 7.65\times 10^{-41} B_{\mu G}^{3.5} ~~~~~({\rm erg\,s^{-1}cm^{-3}Hz^{-1}}).
\label{ReaccEq}
\end{equation}

The observed synchrotron emissivity in the case of radio halos, $\langle J \rangle\simeq 10^{-42}$  erg s$^{-1}$cm$^{-3}$Hz$^{-1}$,
 implies an average magnetic field strength of about 0.3 $\mu$G. At a redshift $z=0.18$, from Eq.\,\ref{Reacc} it follows that the 
required energy supply is $Q_{el}\simeq 2.2\times10^{-30}$  erg s$^{-1}$cm$^{-3}$, about one order of magnitude larger than the required power found in 
the previous section in the case of the continuous injection model. 

A general result from Eq.\,\ref{uelreac} is that there exists a critical value for the magnetic field strength that minimize the acceleration efficiency $\chi=Q_{el}/u_{el}$ needed 
 to radiate at a given $\nu_{max}$, described by  $B= B_{CMB}/\sqrt{3}$. 
In the particular case of the average redshift of halos in our sample, the field that minimizes the re-acceleration efficiency results in $B\simeq 2.6$ $\mu$G. Given this value for the 
magnetic field, from Eq.\,\ref{Reacc} it follows that the required energy supply to the re-acceleration process is $Q_{el}\simeq 3.6\times10^{-32}$  erg s$^{-1}$cm$^{-3}$, i.e. significantly less than in the equipartition case above. However, the energy density of cosmic ray electrons, $u_{el}\simeq 1.3\times 10^{-16}$ erg\,cm$^{-3}$, is much lower than that of the field, $u_{B}\simeq 2.7\times 10^{-13}$ erg\,cm$^{-3}$, and thus the radio source cannot be in a condition of minimum energy.

In Fig.11 we show the comparison of the re-acceleration model versus the exponential fit for the radio halo in A2744. The average spectral index of this radio halo 
is fairly constant with $\alpha_{325 MHz}^{1.4 GHz}\simeq 1$ (Orr\`u et al. 2007) which justifies the use of the same $\nu_{max}$ at different distances from the cluster centre. 
By assuming that $B\propto n_{th}^{\eta}$, from Eq.\,\ref{ReaccEq} it follows that the radio brightness profiles is:
\begin{equation}
I_{1.4}(r)=I_0 (1+r^{2}/r^{2}_c)^{-5.25\beta\eta+0.5}.
\label{profkingreacc}
\end{equation}

If the equipartition energy condition $u_{el}=u_{B}$ is met at any radius, the re-acceleration model yields a central magnetic 
field of $B_0\simeq0.6\, \mu$G with a scaling index of $\eta=0.47$. If we fix $B_{0}=3.2\, \mu$G, we minimize the re-acceleration efficiency 
at the cluster centre but the radio halo is out of equipartition since the total energy density is dominated by the magnetic field energy
 density ($u_{el}\simeq 3\times 10^{-3}u_{B}$). Whatever the case, we find that the exponential profile is also very close to the expectations of the simple re-acceleration model considered here.

\begin{figure*}
\begin{center}
\includegraphics[width=18cm]{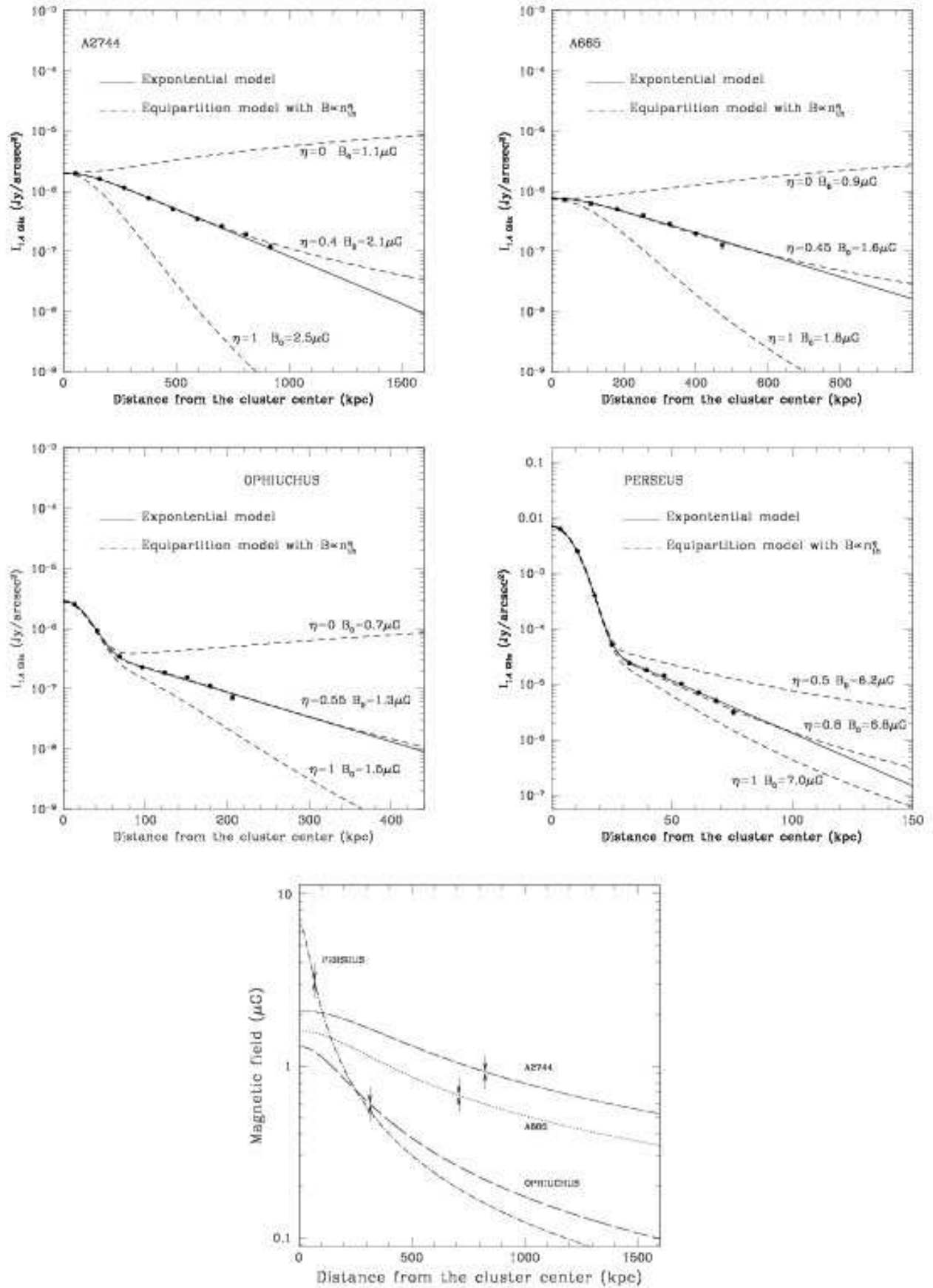}
\caption{Comparison of the exponential (solid line) and equipartition models (dashed lines) with $B\propto n_{th}^{\eta}$ for the radio halos in A2744 and A655 (top panels) and for the mini-halos in Ophiuchus and Perseus clusters (middle panels). The value of the central magnetic 
field strength, $B_0$, is reported along with the value of $\eta$ for each profile. Bottom panel shows the magnetic field trends for the best equipartition profiles. The arrows indicate the
 position of the $3r_{e}$ obtained by the fit using the exponential model.}
\label{fig10}
\end{center}
\end{figure*}

\section{Conclusions}

Mini-halos in clusters are still poorly understood sources.  They are
a rare phenomenon, having been found so far only in few clusters.  A
larger number of mini-halos and better information on their physical
properties will be necessary in order to discriminate between the
different mechanisms suggested for transferring energy to the
relativistic electrons that power the radio emission.

\begin{figure}
\begin{center}
\includegraphics[width=9cm]{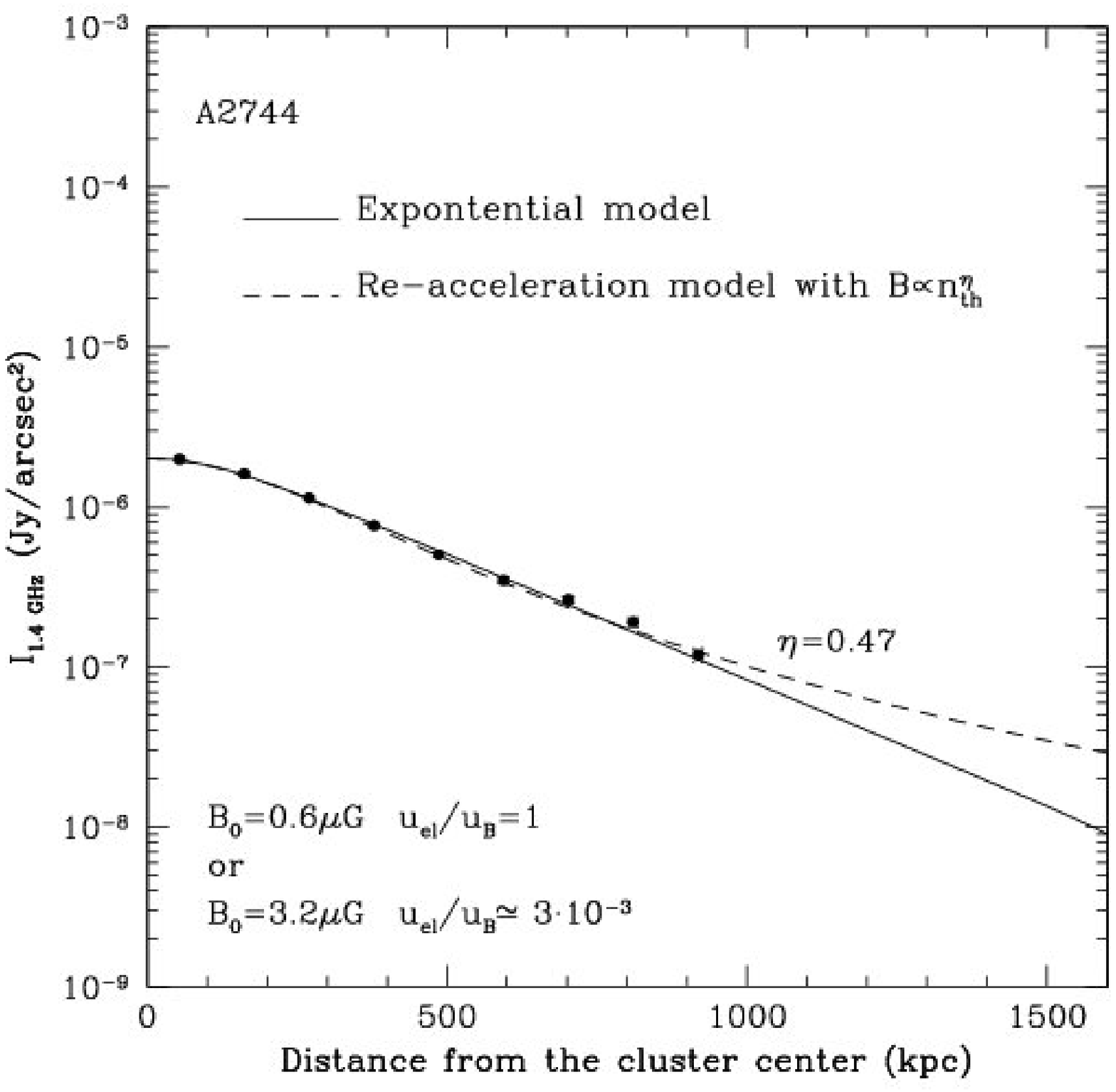}
\caption{Abell 2744. Comparison of the exponential (solid line) and re-acceleration model (dashed line).}
\label{fig11}
\end{center}
\end{figure}

We recently found that at the center of the clusters A1835, A2029, and
Ophiuchus, the dominant radio galaxy is surrounded by a diffuse low
surface brightness mini-halo.  We have studied the morphological and
physical properties (i.e. length-scale, central brightness,
emissivity) of these mini-halos by fitting their azimuthally averaged
radio brightness profile with an exponential.  The method proposed
here to derive the length-scale of halos and mini-halos seems
promising because it is relatively independent of the sensitivity of
the radio observations.  The exponential model is attractive in its
simplicity and involves a minimal set of free parameters.  While it
cannot account for any local deviations from circular symmetry of the
diffuse emission, this empirical method does provide a rough estimate
of brightness and size of these sources.  We compare the surface
brightness profiles of the new mini-halos discovered in Paper I with
data already available in the literature, both for mini-halos and
halos.  We find that the radio halos can have quite different
length-scales but their emissivity is remarkably similar from one halo
to another. This result could have important implications for theories
of the origin of radio halos in clusters of galaxies but awaits
confirmation from future, more sensitive, observations.  In fact, if
many faint halos have been missed by the current surveys, the average
radio emissivity we found at 1.4 GHz, $\langle J \rangle\simeq
10^{-42}$ erg s$^{-1}$cm$^{-3}$Hz$^{-1}$, should be considered an
upper limit. In contrast, mini-halos span a wide range of radio
emissivity. Some of them, like the Perseus mini-halos, are
characterized by a radio emissivity which is more than 100 times
greater than that of radio halos. On the other hand, the new
mini-halos in cooling core clusters analyzed here, namely A2029,
Ophiuchus, and A1835, have a radio emissivity which is much more
typical of halos in merging clusters rather than to the mini-halos
previously known.

We discussed the exponential fit in comparison with the current
theoretical models for the distributions of synchrotron electrons and
magnetic fields in cluster of galaxies.  We find that the exponential
profile is very close to the expectations of these models if the
magnetic field energy density roughly scales as the thermal gas
density. In the framework of the equipartition model, the physical
meaning of the $3r_{e}$ length-scale obtained by the exponential fit
is that it marks the point at which the magnetic field strength is
decreased to about half the value at the cluster center. Mini-halos
would appear smaller because the magnetic field falls off more rapidly
with radius in cooling core clusters, in agreement with the
suggestions of Burns et al. (1992).

  Can we explain the larger dispersion in the emissivity of the
  mini-halos?  We have very bright mini-halos, like Perseus, which
  host a particularly bright compact radio source at their center. For
  these, we cannot rule out that part of the mini-halo radio emission is
  related to properties of the local intergalactic medium and part is
  correlated with the AGN activity of the central brightest galaxy, as
  suggested by the faint correlation, discussed in Paper I, between
  the mini-halo and cD radio power.  Indeed, the higher emissivity
  found in some mini-halos could be due to an extra-amount of energy
  supply from the strong AGN.

  On the other hand, we also observe low-surface brightness
  mini-halos, like Ophiuchus, which host a faint AGN at their
  center. These mini-halos appear to be scaled down versions of the
  larger halos and they could be powered by similar merging
  processes. We note that, although cooling core clusters are
  generally considered relaxed system, when analyzed in detail they
  sometimes reveal peculiar X-ray features in the cluster center which may
  indicate a possible link between the mini-halo emission and some
  minor merger activity. Indeed, Burns et al. (2008) simulated the
  formation of both cool core and non-cool core clusters in the same
  numerical volume.  These simulations confirmed that non-cool
  clusters are formed via major mergers early in their history which
  destroyed the cool cores and left significant residual kinetic
  energy in the gas which might be used to power the radio halos.
  In contrast, cool core clusters do not suffer any major mergers,
  thus preserving the central cool regions.  However, the cool core
  clusters do experience regular smaller mergers which still inject
  energy in the intra-cluster medium, but more modestly than in the
  non-cool core clusters.  These minor mergers might power the
  mini-halos.

\begin{table*}
\caption{Radio information of halos and mini halos taken from the literature
and reanalyzed in this work with the exponential model fit.}
\begin{center}
\begin{tabular} {cclcccccccc} 
\hline
Cluster & Type & Reference           & z     & kpc/$''$ & FWHM &  $I_0$               &  $r_e$      &  $r_e$    &  $\langle J\rangle$ & $\chi^2_{RED}$ \\
        &      &                     &       &          & arcsec & $\mu$Jy/arcsec$^2$    &  arcsec    &  kpc &   erg s$^{-1}$Hz$^{-1}$cm$^{-3}$ &              \\ 
\hline

\vspace{0.2cm}
A2744   &H & Govoni et al. 2001      &0.3080 & 4.50 & 50 & $3.05^{+0.12}_{-0.12}$  & $61^{+2}_{-2}$ & $275^{+9}_{-9}$ & $2.5^{+0.17}_{-0.16}\times 10^{-42}$  & 1.5  \\
\vspace{0.2cm}
A665    &H & Giovannini Feretti 2000 &0.1819 & 3.03 & 53 & $1.09^{+0.08}_{-0.08}$  & $78^{+6}_{-5}$ & $236^{+18}_{-15}$ & $7.0^{+1.0}_{-1.0}\times 10^{-43}$  & 0.9  \\ 
\vspace{0.2cm}
A2219   &H & Bacchi et al. 2003      &0.2256 & 3.59 & 53 & $1.10^{+0.08}_{-0.08}$  & $100^{+5}_{-4}$ & $359^{+18}_{-14}$ & $5.4^{+0.6}_{-0.6}\times 10^{-43}$  &3.7  \\    
\vspace{0.2cm}
A2255   &H & Govoni et al. 2005      &0.0806 & 1.50 & 25 & $0.65^{+0.02}_{-0.02}$ & $135^{+4}_{-4}$ &  $203^{+6}_{-6}$  & $3.4^{+0.20}_{-0.19}\times 10^{-43}$  &  4.8  \\ 
\vspace{0.2cm}
A773    &H & Govoni et al. 2001      &0.2170 & 3.48 & 30 & $0.75^{+0.07}_{-0.07}$  & $32^{+3}_{-3}$ &  $111^{+10}_{-10}$& $1.1^{+0.24}_{-0.2}\times 10^{-42}$  &  0.7 \\ 
\vspace{0.2cm}
A545    &H & Bacchi et al. 2003      &0.1540 & 2.64 & 45 & $1.31^{+0.15}_{-0.13}$  & $57^{+4}_{-4}$ &  $150^{+11}_{-11}$& $1.2^{+0.22}_{-0.19}\times 10^{-42}$  & 0.6   \\ 
\vspace{0.2cm}
A2319   &H & Feretti et al. 1997     &0.0557 & 1.07 & 30 & $1.11^{+0.04}_{-0.04}$  & $185^{+7}_{-6}$ & $198^{+7}_{-6}$ & $5.4^{+0.35}_{-0.36}\times 10^{-43}$  & 2.4   \\ 
\vspace{0.2cm}
A2218   &H & Giovannini Feretti 2000 &0.1756 & 2.94 & 35 & $1.06^{+0.34}_{-0.27}$  & $26^{+9}_{-6}$ &  $76^{+26}_{-18}$& $2.0^{+1.4}_{-0.9}\times 10^{-42}$  &1.0   \\     
\vspace{0.2cm}
A2163   &H & Feretti et al. 2001     &0.2030 & 3.31 & 62 & $2.23^{+0.07}_{-0.07}$  & $119^{+2}_{-2}$ &  $394^{+7}_{-7}$& $9.2^{+0.4}_{-0.4}\times 10^{-43}$  &2.1    \\  
\vspace{0.2cm}
A401    &H & Bacchi et al. 2003      &0.0737 & 1.38 & 45 & $0.44^{+0.06}_{-0.05}$  & $79^{+15}_{-11}$ &  $109^{+21}_{-15}$& $4.1^{+1.3}_{-1.0}\times 10^{-43}$  &0.7   \\  
\vspace{0.2cm}
A2254   &H & Govoni et al. 2001      &0.1780 & 2.98 & 45 &  $1.56^{+0.22}_{-0.20}$  & $80^{+19}_{-13}$ &  $238^{+57}_{-39}$& $9.7^{+3.4}_{-2.8}\times 10^{-43}$  &1.1    \\  
\vspace{0.2cm}
RXJ1314 &H & Feretti et al. 2005     &0.2439 & 3.81 & 45 & $1.05^{+0.31}_{-0.25}$  & $42^{+16}_{-10}$ &  $160^{+61}_{-38}$& $1.2^{+0.8}_{-0.5}\times 10^{-42}$  &0.1    \\  
\vspace{0.2cm}
RXJ1347 &MH& Gitti et al. 2007       &0.4510 & 5.74 & 18 &  $26.0^{+40.7}_{-15.4}$  & $9^{+3}_{-2}$ &  $52^{+17}_{-11}$& $1.8^{+0.7}_{-0.5}\times 10^{-40}$  &0.2  \\ 
\vspace{0.2cm}
A2390   &MH& Bacchi et al. 2003      &0.2280 & 3.62 & 20 & $60.8^{+27.0}_{-21.1}$  & $10^{+1}_{-1}$ &  $ 36^{+4}_{-4} $ & $3.1^{+1.0}_{-0.8}\times 10^{-40}$  &1.2   \\ 
\vspace{0.2cm}
Perseus & MH  & Pedlar et al. 1990   &0.0179 & 0.36 & 45 & $99.6^{+10.9}_{-10.6}$ & $64^{+4}_{-3}$ &  $23^{+1}_{-1}$ &   $3.6^{+0.5}_{-0.5}\times 10^{-40}$  &2.4 \\    
\hline
\multicolumn{8}{l}{\scriptsize Col. 2: Type of diffuse emission contained (H=halo, MH=mini halo);}\\
\multicolumn{8}{l}{\scriptsize Col. 6: FWHM of the circular Gaussian beam;}\\ 
\multicolumn{8}{l}{\scriptsize Col. 10: Average radio emissivity over the volume of a sphere of radius 3$r_{e}$, $k$-corrected with $\alpha=1$.}\\

\end{tabular}
\label{tab3}
\end{center}
\end{table*}

\begin{acknowledgements}

  We are grateful to an anonymous referee for very useful comments
  that improved the paper.  This work is part of the ``Cybersar''
  Project, which is managed by the COSMOLAB Regional Consortium with
  the financial support of the Italian Ministry of University and
  Research (MUR), in the context of the ``Piano Operativo Nazionale
  Ricerca Scientifica, Sviluppo Tecnologico, Alta Formazione (PON
  2000-2006)''.  FG and MM thank the hospitality of the
  Harvard-Smithsonian Center for Astrophysics where most of this
  research was done.  Support was provided by Chandra grants GO5-6123X
  and GO6-7126X, NASA contract NAS8-39073, and the Smithsonian
  Institution. This research was partially supported by ASI-INAF
  I/088/06/0 - High Energy Astrophysics and PRIN-INAF2005.  We wish to
  thank Pasquale Mazzotta for his valuable comments on the original
  draft.  We are also grateful to Rossella Cassano and Chiara Ferrari
  for very useful discussions.  The National Radio Astronomy
  Observatory (NRAO) is a facility of the National Science Foundation,
  operated under cooperative agreement by Associated Universities,
  Inc.  This research has made use of the NASA/IPAC Extragalactic Data
  Base (NED) which is operated by the JPL, California Institute of
  Technology, under contract with the National Aeronautics and Space
  Administration.

\end{acknowledgements}

\end{document}